# ezyMRI: How to build an MRI machine from scratch - Experience from a four-day hackathon

Shaoying Huang,[1] José Miguel Algarín,[2] Joseba Alonso,[3] Anieyrudh R,[4] Jose Borreguero,[5] Fabian Bschorr,[6] Paul Cassidy,[7] Wei Ming Choo,[8] David Corcos,[9] Teresa Guallart-Naval,[10] Heng Jing Han,[11] Kay Chioma Igwe,[12] Jacob Kang,[13] Joe Li,[14] Sebastian Littin,[15] Jie Liu,[16] Gonzalo Gabriel Rodriguez,[17] Eddy Solomon,[18] Li-Kuo Tan,[19] Rui Tian,[20] Andrew Webb,[21] Susanna Weber,[22] Dan Xiao,[23] Minxuan Xu,[24] Wenwei Yu,[25] Zhiyong Zhang,[26] Isabelle Zinghini,[27] Bernhard Blümich[28]

1. Shaoying Huang, Singapore University of Technology and Design, Singapore, huangshaoying@sutd.edu.sg
2. José Miguel Algarín, Instituto de Instrumentación para Imagen Molecular, Valencia, Spain, josalggui@i3m.upv.es
3. Joseba Alonso, Instituto de Instrumentación para Imagen Molecular, Valencia, Spain, joseba.alonso@i3m.upv.es
4. Anieyrudh R, Singapore University of Technology and Design, Singapore, anieyrudh_r@mymail.sutd.edu.sg
5. Jose Borreguero, Instituto de Instrumentación para Imagen Molecular, Valencia, Spain, pepe.morata@i3m.upv.es
6. Fabian Bschorr, Department of Internal Medicine II, University Ulm Medical Center, Ulm, Germany, fabian.bschorr@uni-ulm.de
7. Paul Cassidy, Institute of Molecular and Cell Biology, Agency for Science, Technology and Research (A*STAR), Singapore, paul_cassidy@imcb.a-star.edu.sg
8. Wei Ming Choo, Engineering Product Development, Singapore University of Technology and Design, weiming_choo@mymail.sutd.edu.sg
9. David Corcos, Agile MRI ltd, Tel Aviv, Israel, david.corcos@agilemri.com
10. Teresa Guallart-Naval, Instituto de Instrumentación para Imagen Molecular, Valencia, Spain, tguanav@i3m.upv.es
11. Heng Jing Han, Singapore University of Technology and Design, Singapore, jinghan_heng@mymail.sutd.edu.sg
12. Kay Chioma Igwe, Department of Biomedical Engineering. Columbia University, New York, USA, kci2104@columbia.edu
13. Jacob Kang, Singapore University of Technology and Design, Singapore, jacob_kang@sutd.edu.sg
14. Bing Keong (Joe) Li, Jiangsu LiCi Medical Device Co. Ltd, joeli@licimedical.com
15. Sebastian Littin, Division of Medical Physics, Department of Diagnostic and Interventional Radiology, University Medical Center Freiburg, Faculty of Medicine, University of Freiburg, Freiburg, Germany, sebastian.littin@uniklinik-freiburg.de
16. Jie Liu, NingBo ChuanShanJia Electrical and Mechanical Co., Ltd, bme_lj@163.com
17. Gonzalo Gabriel Rodriguez, Max-Planck Institute for Multidisciplinary Sciences, Göttingen, Germany, gonzalogabriel.rodriguez@mpinat.mpg.de
18. Eddy Solomon, Weill Cornell Medical College, New York, N.Y., USA, eds4001@med.cornell.edu

[19] Li-Kuo Tan, Department of Biomedical Imaging, Faculty of Medicine, Universiti Malaya, Malaysia, lktan@um.edu.my
[20] Rui Tian, High-Field MR center, Max Planck Institute for Biological Cybernetics, Tübingen, Germany, rui.tian@tuebingen.mpg.de
[21] Andrew Webb, Leids Universitair Medisch Centrum, Leiden University, Leiden, The Netherlands, a.webb@lumc.nl
[22] Susanna Weber, Department of Biomedical Engineering. Columbia University, New York, USA, smw2251@columbia.edu
[23] Dan Xiao, Department of Physics, University of Windsor, Windsor, ON, Canada, Dan.Xiao@uwindsor.ca
[24] Minxuan Xu, Department of Electrical and Computer Engineering, College of Design and Engineering, National University of Singapore, minxuan_xu@u.nus.edu
[25] Wenwei Yu, Center for Frontier Medical Engineering, Chiba University, Chiba, Japan, yuwill@faculty.chiba-u.jp
[26] Zhiyong Zhang, School of Biomedical Engineering, Shanghai Jiao Tong University, China, zhiyong.zhang@sjtu.edu.cn
[27] Isabelle Zinghini, Department of Biomedical Engineering, Columbia University, New York, USA, iaz2112@columbia.edu
[28] Bernhard Blümich, Institut für Technische und Makromolekulare Chemie, RWTH Aachen University, Aachen, Germany, bluemich@itmc.rwth-aachen.de

**Graphical abstract**

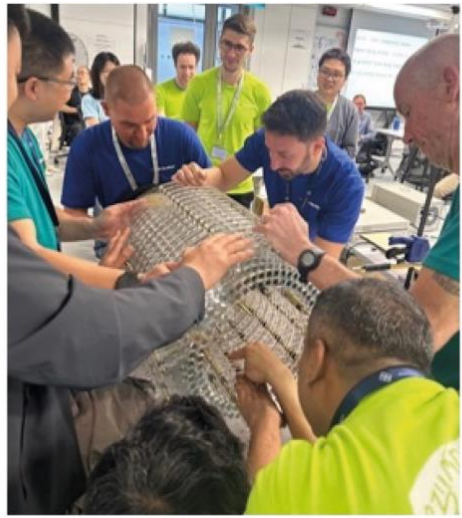

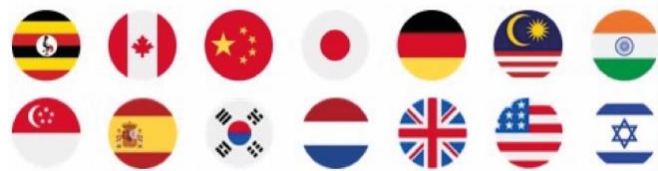

**Keywordstable**

NMR, Low-field MRI, DIY, hackathon

**Abstract**

Nuclear magnetic resonance instruments are becoming available to the do-it-yourself community. The challenges encountered in the endeavor to build a magnetic resonance imaging instrument from scratch were confronted in a four-day hackathon at Singapore University of Technology and Design in spring 2024. One day was devoted to educational lectures and three days to system construction and testing. Seventy young researchers from all parts of the world formed six teams focusing on magnet, gradient coil, RF coil, console, system integration, and design, which together produced a working MRI instrument in three days. The different steps, encountered challenges, and their solutions are reported.

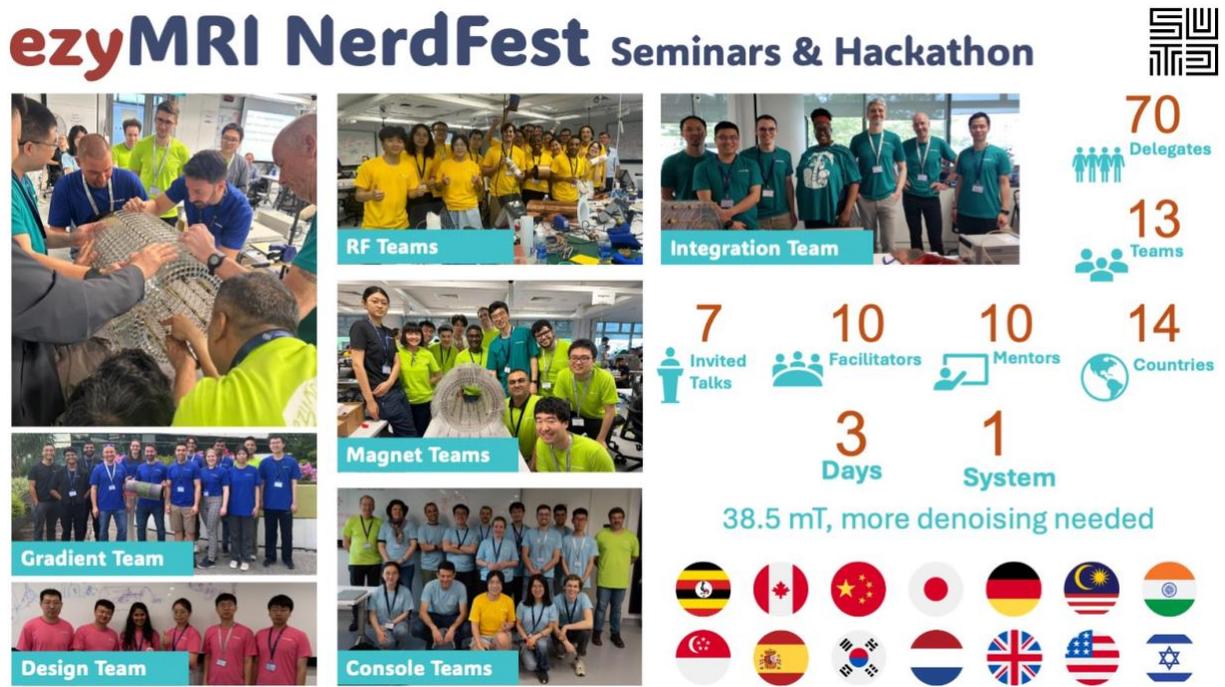

Figure 1. Structure and statistics of the 2024 ezyMRI hackathon at Singapore University of Technology.

## 1. Introduction

In spring 2024 (29 April-3 May 2024), a seminar and hackathon event named ezyMRI NerdFest was held at Singapore University of Technology and Design (SUTD) with the mission of training people to build a low-field magnetic resonance imaging (MRI) instrument from scratch, promoting MRI technology and techniques (especially portable MRI), improving MRI accessibility, and fostering the networking of like-minded engineers and researchers. Seventy delegates from 14 countries participated in the event. Participants first received training through lectures in a one-day workshop and then were divided into six teams led by mentors, each with a different task to produce an MRI machine within 3 days [1]. These teams included the magnet, console, gradient,

RF, integration, and design teams (Fig. 1). This report first introduces the relevant history and characteristics of low-field portable MRI, and then details the work performed at the hackathon.

**1.1 Low-field NMR with permanent magnets**

Instruments with permanent magnets date back to the early days of NMR [2-4]. Because electromagnets are more suited for recording NMR spectra by sweeping the magnetic field at constant radio-frequency (RF) irradiation, permanent magnets were gradually replaced by electromagnets. Nevertheless, the workhorses for chemical analysis from the 1960s onwards were 60 MHz NMR spectrometers with permanent magnets [5] until higher field strengths became available for NMR spectroscopy, with superconducting magnets providing higher spin polarization and wider spectral dispersion [6]. While the size of NMR systems kept increasing with superconducting magnets achieving higher and higher field strength, compact tabletop NMR instruments with permanent magnets became available in the early 1970s for relaxation analysis of foodstuffs and other proton-rich soft matter [7,8]. However, compared to NMR spectroscopy for chemical analysis, NMR relaxometry for materials testing remained a curiosity due to the advent of pulsed NMR, 2D-NMR spectroscopy, and high-field superconducting magnets serving the needs of a much larger community of chemists. Today, most NMR instruments use expensive superconducting magnets with field strengths much higher than the 2.5 T (106.45 MHz) for $^1$H NMR achievable with permanent magnets, such that NMR instruments that operate at $^1$H NMR frequencies of about 100 MHz and lower are often referred to as low-field instruments [5].

Tabletop NMR relaxometers employ permanent magnets where the object is placed in the center of the magnet (i.e., *in-situ*). This is the conventional geometry preferred in most NMR modalities, be it relaxometry, imaging, or spectroscopy. Several alternative specialized geometries, such as NMR well-logging instruments introduced in 1980 for the petroleum industry, have been developed for deployment inside boreholes to investigate fluids in the surrounding wall's pores, with the signal-bearing region located outside the magnet (*ex-situ*) [9]. This modality of stray-field relaxometry can be used not only for well-logging but for a large variety of applications to nondestructive testing of materials [10,11]. As the signal-bearing volume of a stray-field sensor is localized somewhere in the object near the sensor, the information acquired with such sensors is equivalent to that contained in a pixel of a magnetic resonance image. The use of stray-field NMR for materials testing was boosted with the advent of the NMR-MOUSE (MObile Universal Surface Explorer) in 1996 [12] as the capabilities of stray-field NMR in terms of methodology and applications were systematically explored. With time, other permanent magnet designs were and are still being investigated [10,13,14]. In that context the seminal work of Klaus Halbach [15] was brought to the attention of the NMR community in 2004 in a paper describing how to assemble dipolar center-field magnets from many identical magnet cubes [16]. This design is at the origin of the magnet built in the 2024 ezyMRI hackathon [17].

While the early MRI instruments employed conventional electromagnets [18], most imaging systems today rely on superconducting electromagnets, even those at a low field of 0.55 T corresponding to 23.42 MHz $^1$H-NMR frequency, because a stable magnetic field sufficiently homogeneous with a bandwidth usually less than 100 part per million for MRI needs to be maintained for human applications across a diameter of 40 cm or so. Operating clinically useful MRI instruments at even lower fields appeared to be inconceivable for a long time, but are now receiving increasing attention [19].

## 1.2 MRI of materials and processes

MRI is best known for its use and unsurpassed benefits in clinical diagnostics, where it serves to reveal the morphology and function of soft-matter tissue with rich contrast [20-22]. While biomedical studies are at the forefront of MRI research [21,22], MRI also has a long tradition of use in chemical engineering, agriculture, geology, and materials science to study the properties and function of materials and technical devices in a much smaller community [23-27]. Areas of investigation are, for example, the flooding and displacement of fluids in porous rocks for the petrochemical industry, moisture transport and associated phenomena in building materials, the fluid transport and reaction in model and micro-reactors, the rheology of non-Newtonian fluids including granular matter, the dissolution of tablets and drying of paint, the distribution of mechanical properties in composite rubber products like tires, and the growth of plants along with the quality control of agricultural products [27-32]. Considerable progress in measurement methodology and instrumentation came out of this community, such as dedicated scanners for rapid screening of oranges [33,34] as well as mobile instruments for MRI of human limbs and fruit growing on trees [35].

## 1.3 The need for and use of affordable MRI for medical diagnostics

MRI is an indispensable part of clinical care, with over 100 million scans performed worldwide annually and around 50,000 machines in hospitals and clinics. However, people in low- and middle-income countries (LMICs) have severely limited access to this technology, despite constituting more than 70% of the world's population [36-43]. For example, there are 0.7 MRI scanners reported per million people in Africa and far fewer in countries such as Mali and Niger, compared to 55 in Japan, 40 in the USA and 35 in Germany [43]. Most are in major cities, far from rural populations. Of the scanners in Africa, 39% use obsolete hardware and software.

Clinical MRI is expensive. Usually, it takes several millions of dollars to purchase, install, maintain, and operate one machine. Moreover, clinical MRI is bulky, requiring a superconducting magnet several meters in size, weighing 5–10 tonnes, cooled by liquid helium, and housed in a shielded room to reduce electromagnetic interference. Clinical MRI needs high electrical power and chilled water. Highly trained technicians are also essential to operate the machine. Software and hardware upgrades cost tens to hundreds of thousands of dollars each year. All this is beyond the means of healthcare, research, and education settings in LMICs. Instead, LMICs need completely redesigned MRI systems that can be set up and run easily in rural settings and secondary health facilities. Local physicians need to have confidence in such redesigned forms of MRI to diagnose and treat common life-threatening conditions.

Today, most clinical MRI scanners operate at high magnetic fields (1.5 T – 3 T), and their use is mostly limited to diagnostic purposes rather than a first-line screening modality due to high cost and limited accessibility. As an alternative, low-field MRI scanners are becoming increasingly attractive and accessible for clinical use due to the reduced cost and the reduced size/footprint [44]. In a hospital environment, lower-field scanners can reduce the cost of the magnet and lower maintenance expenses. The utility of low-field magnets can become applicable either by ramping down commercial magnets [44], or by exploring new dedicated lower-field machines. Their use also holds great potential for overcoming some of the basic limitations of today's high-field clinical MRI magnets, such as allowing wider bore configurations, relaxed safety concerns for metallic implants, and decreased specific absorption rate (SAR) [45]. These advances can increase accessibility for patients who would typically not qualify for a routine MRI scan.

Recent studies demonstrate the potential use of low-field systems in multiple body applications such as functional cardiac imaging [46], pulmonary imaging [47], abdominal imaging [48], and musculoskeletal imaging [49]. In many of these studies, a high-performance low-field MRI system (0.55 T, 45 mT/m) [45] could provide comparable and sometimes even better results than high-field instruments. Nevertheless, addressing issues like reduced gradient performance, spatial field inhomogeneity, and prolonged scanning times remains an active area of research. Finding solutions to these challenges would enable the use of single-shot rapid imaging sequences that are known to be susceptible to local field inhomogeneities, such as echo-planar imaging (EPI) [50,51].

Today, permanent magnets are gaining traction in all modalities of NMR because they do not need to be serviced with electric energy or cryo-coolants to maintain their field strength, and they can be assembled at comparatively low cost [52]. Notably, it has been demonstrated that medically pertinent magnetic resonance images of the human brain can be generated at magnetic fields of 0.1 T (4 MHz) and less, and that image quality can greatly be improved with tools from machine learning and artificial intelligence [53,54]. Even lower fields are conceivable for MRI if the image-acquisition process is assisted by hyperpolarization strategies [55-57] and/or ultrasensitive detection [58,59]. Low-field MRI with permanent magnets has matured from affordable laboratory equipment for research and education [60] to the stage where the entire instrument is being assembled and tested in Africa in a point-of-care setting [39]. The ezyMRI hackathon's aim was to reproduce the construction of a low-field MRI instrument, given the open-source information available in the literature and on the internet. The use of permanent magnets, known for their low cost and availability, made the hackathon feasible and facilitated the promotion of low-cost MRI technology and techniques.

### 1.4 Essential MRI

In NMR transverse nuclear magnetization $M_{xy}$ is measured, which in the simplest case has been generated from magnetic polarization of nuclear spins exposed to a static magnetic field ***B*** following a radio-frequency (RF) pulse with a magnetic-field orientation perpendicular to ***B***. This magnetization precesses (rotates) around the direction of ***B*** with an angular frequency $\omega = \gamma\, B$, where $B$ is the magnitude of the magnetic field vector $\mathbf{B}$, and $\gamma$ is the gyromagnetic ratio of the chosen nucleus.

Unlike other spectroscopic methods such as infrared spectroscopy, X-ray scattering, or neutron scattering, NMR is unique in that the phase $\varphi = \omega\, t = \gamma\, B\, t$ of the precessing magnetization vector can be measured,

$$M_{xy} = M \exp\{i\varphi\} = M \exp\{i\gamma\, B\, t\}, \tag{1}$$

where $t$ is the observation time [25,60], and $M$ is the projection of the magnetization in the transverse plane. Note, that both the magnetization amplitude $M$ and the precession frequency $\omega$ are proportional to the static magnetic field $B$. Therefore, higher magnetic fields have traditionally been pursued to increase the signal-to-noise ratio (SNR).

In a linearly varying magnetic field $\mathbf{B} = \mathbf{B}_0 + \mathbf{G}\, \mathbf{r}$ with a constant gradient $\mathbf{G} = [G_x, G_y, G_z]^t$, the precession phase varies linearly with the position $\mathbf{r} = [x, y, z]^t$ of the magnetization, such that

$$M_{xy} = M \exp\{i\gamma\, (\mathbf{B}_0 + \mathbf{G}\, \mathbf{r})\, t\} = M \exp\{i\gamma\, B_0\, t\} \exp\{i\gamma\, \mathbf{G}\, \mathbf{r}\, t\}. \tag{2}$$

Strictly speaking, the gradient is a tensor with nine components, but in a strong polarization field $\mathbf{B}_0$, where the gradient field $\mathbf{G}\, \mathbf{r}$ is small and parallel to $\mathbf{B}_0$, the three components associated with the direction of the polarization field dominate, and the other six components can be neglected, such that eqn. (2) is a valid approximation.

The equation formally simplifies when it is written in a coordinate frame that rotates with the offset frequency $\omega_0 = \gamma\, B_0$,

$$M'_{xy} = M_{xy} \exp\{-i\, \omega_0\, t\,\} = M \exp\{i\gamma\, \mathbf{G}\, \mathbf{r}\, t\}. \tag{3}$$

When the magnetization moves during the observation time, i. e. $\mathbf{r} = \mathbf{r}(t)$, for example, if the molecules are displaced by diffusion or flow, the magnetization experiences a time varying magnetic field $B(t) = B_0 + \mathbf{G}\, \mathbf{r}(t)$, and the phase is determined by the time integral of $B(t)$, such that

$$M'_{xy} = M \exp\{i\, \gamma\, \textstyle\int_0^t \mathbf{G}\, \mathbf{r}(t)\, dt\}. \tag{4}$$

To explain how different characteristics such as diffusion and flow can be encoded into an image, it is helpful to expand the time dependence of the space vector into a Taylor series, $\mathbf{r}(t) = \mathbf{r}_0 + \mathbf{v}_0\, t + \tfrac{1}{2}\, \mathbf{a}_0\, t^2 + \ldots$ , where $\mathbf{r}_0$, $\mathbf{v}_0$, and $\mathbf{a}_0$ are initial position, velocity, and acceleration, respectively. Furthermore, in MRI, the gradients are typically pulsed, meaning that $\mathbf{G}$ depends on time as determined by the chosen imaging sequence. This leads to the expression of magnetization at one initial position as

$$\begin{aligned}M'_{xy} = {}& M \exp\{i\, \gamma\, \textstyle\int_0^t \mathbf{G}(t)\, t^0\, dt\, \mathbf{r}_0\} \\ & \times \exp\{i\, \gamma\, \textstyle\int_0^t \mathbf{G}(t)\, t^1\, dt\, \mathbf{v}_0\} \\ & \times \exp\{i\, \gamma\, \textstyle\int_0^t \mathbf{G}(t)\, t^2\, dt\, \tfrac{1}{2}\, \mathbf{a}_0\} \times \ldots ,\end{aligned} \tag{5}$$

which contains the different moments $\boldsymbol{\mu}_n = \textstyle\int_0^t \mathbf{G}\, t^n\, dt$ of the gradient modulation function $\mathbf{G}(t)$.

Depending on what is to be measured, i. e. an image of position $\mathbf{r}_0$, a distribution of velocities $\mathbf{v}_0$ or acceleration $\mathbf{a}_0$, the associated moments $\boldsymbol{\mu}_0$, $\boldsymbol{\mu}_1$, and $\boldsymbol{\mu}_2$, respectively, need to be varied in the MRI experiment. If there is no movement within the object, $v_0 = 0 = a_0$, and the signal is simply given by

$$M_{xy} = M \exp\{i \gamma_0 \int^t G(t) \, dt \, r_0\} = M \exp\{i \, k \, r_0\}, \tag{6}$$

where $k = \gamma_0 \int^t G(t) \, dt$ is the wave vector. If there is movement, a gradient modulation function can be implemented by the pulse sequence to eliminate its first-order moment $\mu_1$ to suppress motion artifacts.

This expression applies to the magnetization at one initial position or pixel $r_0$. The acquired signal $s(k)$ is the sum of all contributions $M_{xy}$ from individual positions $r_0$,

$$s(k) = \iiint M_{xy}(r_0) \, dr_0 = \iiint M(r_0) \exp\{i \, k \, r_0\} \, dr_0. \tag{7}$$

Consequently, the image $M(r_0)$ can be retrieved from the measured $k$-space signal $s(k)$ by Fourier transformation over the wave vector $k$,

$$M(r) = 1/(2\pi)^3 \iiint s(k) \exp\{-i \, k \, r\} \, dk, \tag{8}$$

provided the signal $s(k)$ has been sampled appropriately for all relevant $k$ values, i.e. the different moments $\mu_n$ of the gradient modulation function $G(t)$ have been varied accordingly in the MRI pulse sequence.

The signal $s(k)$ and image $M(r)$ are discretized based on the Nyquist criterion, where the desired image resolution and field of view define the range and intervals of $k$-space that must be sampled. Given a certain signal lifetime, the maximum $k$ value is generally restricted by the maximum gradient amplitude and slew rate. Consequently, in addition to SNR, image resolution is also limited by the gradient hardware. In this brief explanation of MRI, relaxation and the discussion of image contrast have been omitted.

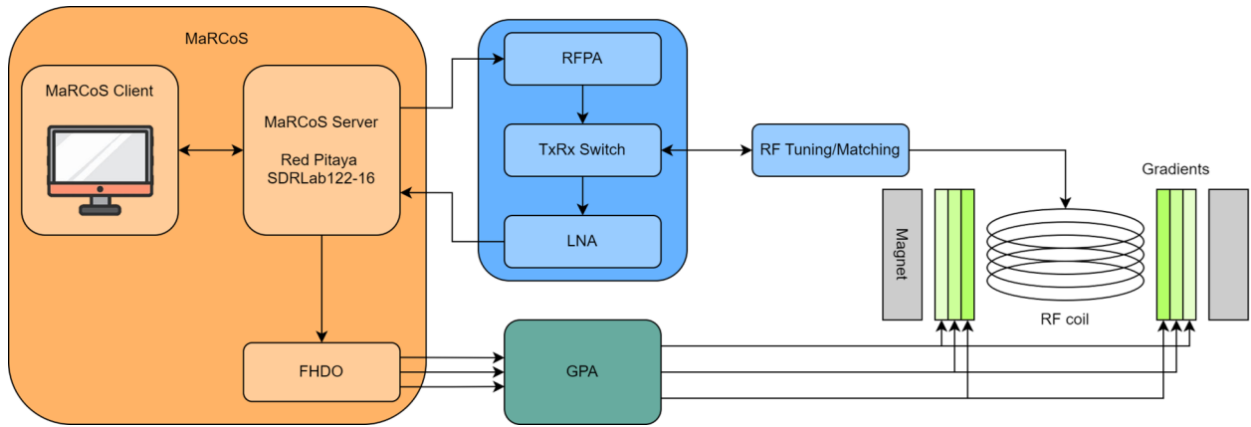

Figure 2. Main components in an MRI system.

The hardware to produce NMR images following equations (5) to (8) consists of the following components (Fig. 2) [61-64]:
- A magnet that provides the polarization field $B_0$,
- an electromagnet that provides pulsed gradient fields $G \, r$ with gradients in $x$, $y$ and $z$ directions,

- gradient power amplifiers (GPA) to enable sufficiently strong pulsed gradient fields,
- a radio-frequency transmitter (Tx) to generate RF pulses,
- an RF coil which acts as an antenna to transmit the RF pulses to the object under study and receives the signal induced by the precessing nuclear magnetization,
- a receiver (Rx) to process the RF signal induced in the coil, and
- a control system or a computer to oversee the MRI pulse sequence and process the data.

All components contribute to the image quality. Quantities such as $B_0$ inhomogeneity and gradient-field nonlinearity must be considered in signal modeling or corrected during image reconstruction. Noise characteristics in the RF chain can significantly affect the image quality. Additionally, the MRI pulse sequence may be limited by the control system. A well-designed MRI scanner strives to achieve high performance in all aspects. These components are further detailed in the subsequent sections.

## 2. Open-source resources

There is an abundance of open-source articles and repositories covering the various aspects of constructing and operating NMR and MRI scanners. Many are included in the initiative *Open Source Imaging* [65] and a paper by its promoters [66]. The potentially useful resources listed in the next three subsections distinguish among open-source hardware designs for the main scanner components, resources focusing on the console and control electronics, and open-source simulators for permanent magnet arrays. It should be noted that, despite the copious information for DIY system design and construction contained within these sources, assembling, fine-tuning, and operating a magnetic resonance scanner is non-trivial, and the overall performance achieved typically depends on the background, skills, and experience of the people involved.

### 2.1 Scanner hardware

Three open-source projects devoted to designing and constructing low-field MRI hardware for educational, research, and clinical purposes are highlighted:
- A low-cost (< 10 kUSD) tabletop scanner was developed at MIT and MGH for educational purposes [60]. This is based on a yoked (H-shaped) permanent magnet generating a field of 180 mT, where samples can be imaged inside 10 mm diameter tubes. Around twenty such systems were used by over 800 students between 2014 and 2020. A wiki including instructions on how to build the system is publicly available [67].
- The Zeugmatron Z1 is the system assembled during the MRI4All hackathon held in New York City in 2023 [68]. It is based on a 44 mT Halbach-array magnet and features a bore opening of 12 cm in diameter, which suffices for some *in vivo* extremity applications. The public repository for the project [69] contains an exhaustive list of the resources and instructions required to build the scanner.
- The OSI2 system [70] is meant as a prospective clinical scanner for accessible neuroimaging. It is largely based on the design of the Leiden University Medical Center [71], which provides a field of 46 mT and features a clear bore opening of 30 cm in diameter.

## 2.2 Console and control electronics

A core functionality of an MRI control system is to handle synchronous pulse sequence execution. This can be realized with Field Programmable Gate Arrays (FPGAs), which are currently inexpensive and widely accessible. Consequently, there are several different solutions available, some commercial, such as those from Pure Devices (Rimpar, Germany), Magritek (Wellington, New Zealand), or Niumag (Suzhou, China), and some publicly available [72-77]. Here two main resources are highlighted.
- The Open-source Console for Real-time Acquisition (OCRA) [78] was first developed for the educational tabletop system from MIT and MGH and is based on a Red Pitaya board. Online documentation is available on GitHub [79].
- MaRCoS (Magnetic Resonance Control System) is an evolution of OCRA, and has been used for this project (Sec. 5).

An important part of the console is the user interface. Presently, there are two different graphical user interfaces (GUIs) for MaRCoS, both written in Python: one developed during the MRI4All hackathon [68] and the other named MaRGE (MaRCoS Graphical Environment) [80]. MaRGE was chosen for this project (Sec. 5.1).

## 2.3 Simulators for permanent magnet arrays (PMAs)

There are open-source simulators to calculate the magnetic field and force of PMAs.
- MagTetris is a fast Python/MatLab-based simulator that can perform fast calculations for both the *B*-field and the magnetic force simultaneously for a PMA consisting of cuboid and arc-shape magnets [81,82].
- Magpylib is a Python-based freeware based on the current model [83], which can handle permanent magnets of cuboid, cylindrical, spherical, and triangle shaped segments with homogeneous magnetization, i.e, magnetization in a single direction and not radial [84].

Employing simulators is critical for the designs of high-performance PMAs.

## 3. Magnet

### 3.1 Magnet designs for low-field MRI

The low cost associated with low-field technology has facilitated the exploration of a diverse range of scanner designs [85]. Regarding the generation of the main magnetic field $B_0$, two main alternatives have been predominantly adopted. One is the implementation of resistive electromagnets, excluding superconducting magnets, and the other is using permanent magnets.

Additionally, there are hybrid systems that combine both resistive electromagnets and permanent magnets [86].

### 3.1.1 Resistive Electromagnets

Resistive electromagnets offer great versatility because the magnetic field strength is controlled through electric current. Thus, with an appropriate control system and power supply, it is possible to gain high robustness to frequency drifts due to temperature variations. Additionally, the capability of changing the field strengths can be exploited to realize experiments that are not possible at fixed fields, such as dispersion-contrast images, proton double irradiation, and proton-electron double resonance images [87-89]. Nevertheless, the necessity of a power supply and, in most cases, a cooling system represents a complication that is a limiting factor for portability. Some recent examples of low-field scanners based on electromagnets include magnets with bi-planar geometry operating at 6.5 mT [90], bi-planar extremity-only magnets [91], and field-cycling scanners, which are capable of rapidly changing the magnetic field from the Earth's magnetic field (or lower) to values exceeding 100 mT in a few milliseconds [92-94].

### 3.1.2 Permanent magnets

In the case of scanners based on permanent magnets, the materials most employed are samarium–cobalt (SmCo) and neodymium-iron-boron (NdFeB). SmCo magnets exhibit a lower temperature remanence coefficient (0.0305%/°C - 0.035%/°C) than NdFeB magnets (0.125%/°C) [91,92]. However, NdFeB offers a higher maximum energy product ($BH)_{max}$ of 35 to 52 MGOe (mega-gauss-oersted), corresponding to the grade of N54, compared to 22 MGOe for SmCo [95,96]. The most common designs are the bi-planar configuration and the Halbach array described below, although there is a wide range of other designs, including single-sided magnets [97,98] and inward-outward ring-pair magnet arrays with incorporated readout gradients [99]. These designs are typically lightweight and open, but the main magnetic field often exhibits significant inhomogeneity along one of the encoding directions, imposing a significant constraint on *k*-space sampling. A review on permanent magnet arrays for portable MRI can be found in the literature [100].

*Bi-planar configuration*
This configuration consists of two magnets (typically disks) placed parallel to one or two ferromagnetic yokes (C- or H-shape magnets) [100]. The yoke increases the main magnetic field and reduces the fringe fields. Typically, additional ferromagnetic rings are used for passive shimming. Most bi-planar designs found in the literature employ SmCo magnets to enhance thermal stability. However, there are also systems that utilize NdFeB magnets [101]. This design usually provides $B_0$ values between 50 mT and 64 mT, with inhomogeneities in the order of 250 ppm (after passive shimming) within a spherical diameter of 20 cm, and the weight of the magnets ranges from 350 to 1,300 kg [101-104]. However, a higher field strength of 110 mT has also been realized for portable MRI [105]. The weight and the strong magnetic force between the magnets require specific machinery that complicates in-house assembling.

*Halbach array*

A Halbach array is an array of permanent magnets arranged in a specific manner that enhances the magnetic field on one side of the array and minimizes the field on the opposite side [15]. The dipolar Halbach array is employed in the design of both NMR and MRI magnets, where permanent magnets are positioned on a series of rings or mandalas [16,106]. This configuration enables the creation of strong fields concentrated within the bore, while simultaneously reducing the intensity of fringe fields. Typically, several thousand small magnets (approximately $12^3$ mm$^3$) are used for the construction of MRI magnets. Most designs employ NdFeB magnets due to their high magnetic field and low cost. Several designs have been successfully implemented for the human brain and extremities images with main magnetic fields ranging from 50 mT to 80 mT, weights from 75 kg to 250 kg, and inhomogeneities from 2000 ppm to 3000 ppm [71,107,108]. When compared to bi-planar configurations, Halbach arrays offer lighter designs with smaller fringe fields and higher inhomogeneity. Furthermore, using multiple smaller magnets instead of just two larger ones reduces the force between them, thereby improving safety and eliminating the necessity for the use of mechanical tools during the assembly process. This enables the construction process to be completed in-house, which was one of the objectives of the ezyMRI hackathon. Additionally, the Halbach array configuration represents the optimal solution for constructing a magnet within a three-day timeframe. The downside of using multiple small magnets is the increase of the total magnet surface area which makes the array more vulnerable to a temperature change compared to one that consists of bigger magnets.

## 3.2 Design process and tools for permanent magnets

### 3.2.1 Design specifications

The first step in the design process is to analyze the specifications of the magnet design and the given constraints. These specifications are typically derived from the requirements of the MRI scanner being developed. They include:
- Effective diameter of spherical volume (DSV), field of view (FoV), and image resolution
- Bore shape and dimensions
- Configuration model (e.g., Halbach type, inward-outward (IO) ring-pair type, H-shape, C-shape, single-sided, etc.) [100]
- Field characteristics (homogeneous or inhomogeneous, presence of built-in gradient, field strength, homogeneity, bandwidth, and 5-gauss zone)
- Footprint dimensions and total system weight

Certain parameters, such as footprint dimensions, the 5-gauss zone, and total weight, may serve as either strict design constraints or flexible design considerations depending on the project.

A given configuration model can be achieved using either a few large permanent magnet pieces (2-5) or a permanent magnet array (PMA) composed of multiple smaller magnets. The former approach generally allows for higher magnetic field strength and homogeneity with the same total magnet weight. However, it requires advanced manufacturing expertise and tooling. The PMA approach, by contrast, offers greater flexibility in terms of design, assembly, and maintenance. Considering the ezyMRI hackathon's do-it-yourself (DIY) spirit, the PMA method was chosen for this project.

In the design of the magnet array, critical parameters include the geometric and magnetic properties of the magnet blocks, specifically, their shape, dimensions, remanence ($B_r$, which is the magnetism that remains after the external magnetic field that magnetized it has been removed), and coercivity ($H_c$, which represents resistance to demagnetization) [100]. The magnet and design specifications for the ezyMRI hackathon are outlined in Table 1.

Table 1. Design parameters of the ezyMRI hackathon magnet

| **Magnets** | | **Image** | |
|---|---|---|---|
| Material | N52 Neodymium | FoV (mm$^3$) | 150 mm DSV |
| Shape | Cubic | | |
| Dimension (mm$^3$) | 12 x 12 x 12 | **Array** | |
| Remanence (T) | 1.43 | Shape | Cylindrical |
| Coercivity (A/m) | 800 - 950 kA/m | Configuration | Halbach |
| | | Dimensions (mm) | Diameter: 220 Length: 52 |

It is important to note that the housing contributes significantly to the overall weight of the system, particularly for the PMA approach. The housing must secure the precise position and orientation of each magnetic block while withstanding the forces and torques generated by the magnetic field. The housing design is discussed in detail below. Although not reflected in the hackathon prototype, factors such as environmental conditions, ergonomics, aesthetics, and maintenance should be considered in a full-scale design.

### 3.2.2 Simulation software for calculating the fields of permanent magnet arrays

Magnetic simulation is usually the same as a low-frequency electromagnetic simulation but based only on the magnetic field without the electric part of the field taken into account. Magnetic simulations are often divided into alternating current (AC), direct current (DC), and transient magnetics. The DC magnetic (magnetostatic) module may be used to analyze problems that require calculating the static magnetic fields caused by a combination of local and/or distributed DC, permanent magnets, and external fields defined by boundary conditions. The DC magnetic module is used to design and analyze devices such as solenoids, electric motors, magnetic shields, permanent magnets, and disk drives [109]. One of the most effective technologies for magnetic simulations is the finite element method (FEM). Commonly used FEM-based electromagnetic field (EM) analysis software can be found in the literature [110].

The PMA approach has a high degree of design flexibility, although in the design cases with many magnetic blocks, the problem represents an enormous search space. Since most optimization methods rely on a forward model to evaluate and update the current solution for a better outcome, effectiveness and efficiency of optimization depend on a fast and accurate magnetic field simulation. However, most FEM-based EM software is computationally intensive if used to calculate the magnetic field of PMA's. Either the solution space needs to be limited, or the accuracy needs to be traded with computation time [107], so optimization with FEM is impractical. Thus, to leverage the design flexibility of PMAs, it is preferable to have a fast and accurate calculation of the magnetic field and magnetic forces of the PMA with sufficient flexibility with respect to the material, shapes, and dimensions of magnets.

MagTetris [81] is an open-source magnet simulator that can satisfy all the requirements for PMA designs. Its accuracy has been verified with respect to the commercial software packages CST STUDIO SUITE® (SIMULIA) and COMSOL Multiphysics® while showing at least 500 times higher computational efficiency than those commercialized FEM models [17]. In this hackathon event, MagTetris was used to design and optimize a Halbach PMA with the design specifications listed in Table 1. The performance parameters of the designed PMA are reported in Table 2. The magnet's bandwidth was calculated by taking the ratio of $B_{max}$ - $B_{min}$ and the average of the magnetic field in the FoV.

Table 2. Performance specifications of the ezyMRI hackathon magnet and scanner

| **Field** | | | |
|---|---|---|---|
| Strength (mT) | 35.2 | Homogeneity (ppm) | 870 |
| Bandwidth | 0.087% | 5 Gauss zone (mm$^3$) | 650 × 870 × 1160 |
| **Scanner** | | | |
| Footprint (mm$^3$) | 460 x 480 x 520 | Total weight (kg) | 66.77 |

### 3.2.3 Magnet-housing design

The magnet housing is a mechanical structure designed to hold the permanent magnet blocks in their correct spatial positions and orientations as specified in the design. It serves multiple purposes: enabling safe transportation, maintaining alignment, protecting the magnets from external impacts, and enabling passive shimming [111,112]. The housing structure and manufacturing method must be tailored to the specific PMA design. Common housing designs include layered [17,113] and integrated structures [98]. For layered structures, laser cutting can be employed on acrylic boards and water jetting on polycarbonate boards. For arrays that consist of magnet cuboids, layered structures usually only have two-dimensional flexibility for the orientation of magnets whereas integrated structures may offer three-dimensional flexibility. The material chosen for the housing must provide the necessary strength while being suitable for forming into the required geometry. Other key considerations include dimensional tolerances (i.e., the acceptable range of variation for specific dimensions) and the ease of assembly.

For this hackathon the housing utilized a layered structure without a fixture base and with long through bolts fastening and aligning all layers together. Figure 3 shows the 3D and front views of the housing. Each layer consists of a 12 mm gap and two 2 mm thick acrylic boards (Fig. 3a). Shorter bolts were used to secure the layers in groups of two with a gap size of 18 mm. The 12 mm thick boards have $(12~\text{mm})^3$ etched voids to house the magnets and the 2 mm ones are fastened to the two sides of the 12 mm boards using short M6 bolts with nuts to secure the positions of the magnets in the longitudinal direction. The length of the short bolts crosses two layers. Besides fastening the three boards together for each layer, the short bolts are used to further fasten the layers. Half of the bolts start from odd number layers and the other half start from even number layers. In the azimuth direction, this scheme is reversed. By doing so, the number of bolts is minimized to reduce the total weight of the PMA. 3D printed cylindrical spacers were hooked onto the screws to keep the required distance between the layers. Details of the magnet housing are provided in Table 3.

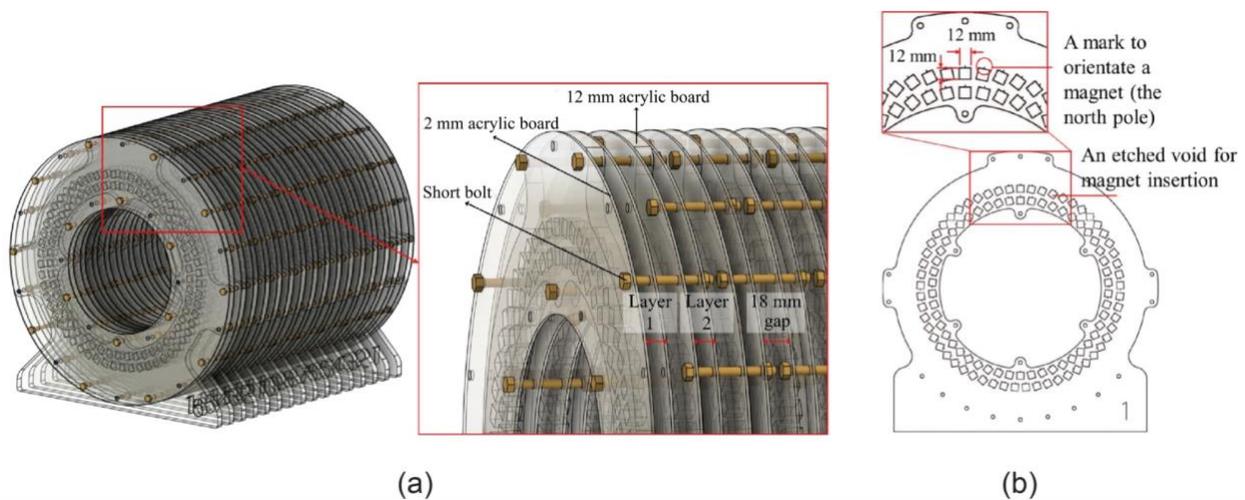

Figure 3. CAD drawing of the housing design. (a) 3D view. (b) Front view.

Table 3. Properties of the magnet housing

| Item | Property |
| --- | --- |
| Housing material | Acrylic |
| Housing weight (kg) | 61.64 |
| Total weight of assembly (kg) | 66.77 |
| Number of layers | 16 |
| Number of long through bolts | 4 |
| ● Material | Brass |
| ● Dimension | M6 × 550 mm |
| Number of short bolts | 210 |
| ● Material | Brass |
| ● Dimension | M6 × 60 mm |
| Weight of bolts and nuts (kg) | 5.13 |

### 3.3 Assembling the magnet

The housing design in the ezyMRI hackathon does not have a fixture base. It requires the magnet layers to be assembled consequentially. For each layer, each (12 mm)$^3$ magnet was inserted to the void after checking its polarization using a polar identification pen. At the edge of a void, a mark was laser cut to indicate the north of a magnet block. A wooden tool was employed to separate individual magnets from a stack of magnets. With this approach, the force to insert a magnet block increases, as more layers are being finished. At the same time this approach does not allow the layers to be assembled simultaneously, which slows down the process. Considering the housing design, a suitable fixture would allow multiple magnet layers to be assembled at the same time. The downside of this other approach is that the fixture will add weight to the PMA.

The magnetic field of the Halbach array was mapped using a two-inch three-axes magnetometer from Lakeshore MMZ-2502-UH. If unavailable, a DIY three-axes magnetometer can be built inexpensively for less than € 200 [114]. A three-axes positioning device was constructed and programmed with MATLAB to automatically map a three-dimensional volume of the magnet. Based on those measurements, the resonance frequency $f_0$ was determined, and all hardware components were tuned accordingly.

### 3.4 Shimming the magnet

Sparse Halbach arrays often experience assembly-induced magnetic field inhomogeneity, which can compromise imaging performance. To mitigate this, shim magnets are employed to correct field distortions and ensure reliable system operation. In the ezyMRI hackathon, a 3D-printing-enabled shimming technique was introduced that allows precise population of shim magnets in the assembled PMA without requiring structural modifications to the array.

3D-printed shim racks were designed to fit seamlessly into pre-existing slots between magnet layers (Fig. 4a-c). There can be multiple ring sectors (Fig. 4a,b). The racks have precisely engineered voids for accurate placements of the shim magnets and custom connectors to secure integration with the existing metal framework (Fig. 4c). The connectors ensure easy installation and robust attachment. This shimming approach eliminates the need for any alteration to the structure of the array.

The complete shimming workflow is outlined in Figure 5. To optimize the magnetic field distribution, a fast genetic algorithm was applied in conjunction with the in-house magnetic field calculator MagTetris [81]. The optimized shim-magnet configuration is shown in Figure 6, where six 3D-printed shim racks were installed between layers 1-2 and 2-3. Figure 7 shows the CAD drawing and photos of the PMA with the shim racks installed. After shimming the measured field distribution of the PMA was compared to the targeted field distribution (Fig. 8). The post-installation measurement revealed a reduction in field inhomogeneity from 194 ppm to 171 ppm, along with a slight increase in average field strength from 38.32 mT to 38.42 mT. Table 4 summarizes the performance parameters of the PMA before and after shimming.

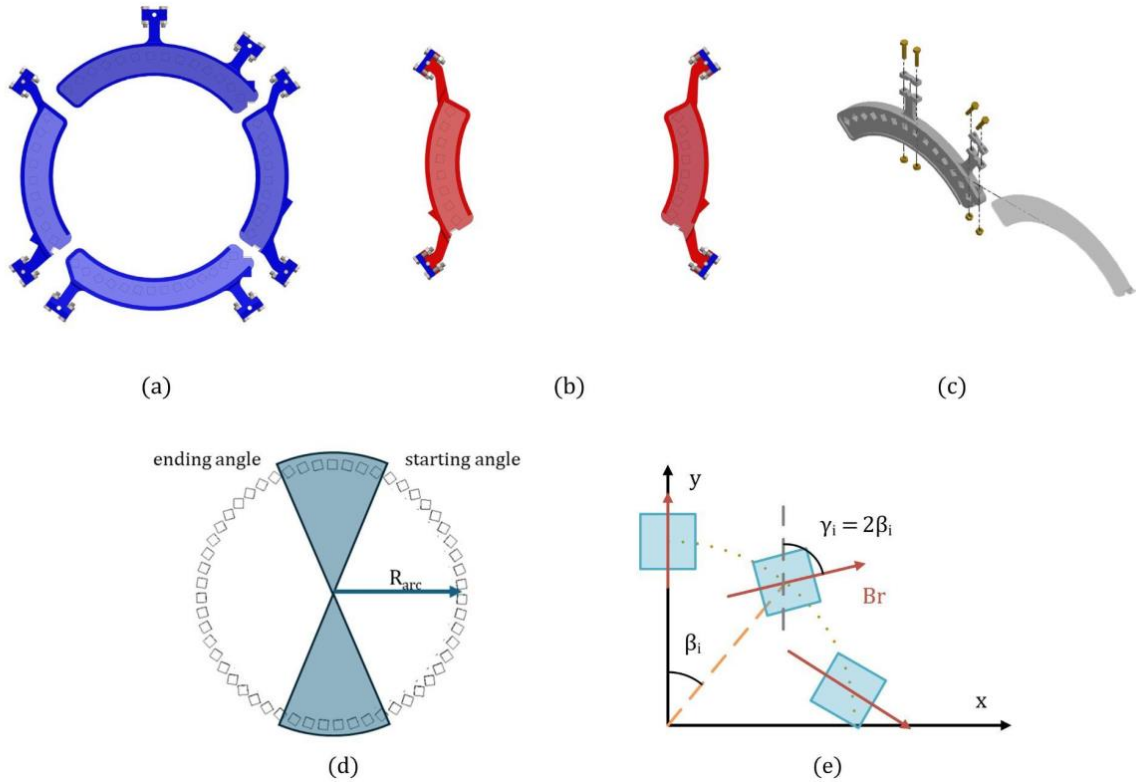

Figure 4. 3D-printed shim-magnet rack. The shim rack is designed to be installed in the target magnet without modifying its original structure. The geometry is highly customizable, featuring a symmetric arc shape that adheres to the Halbach distribution but allows flexible selection of specific segments of the circle, enabling more precise fine-tuning. (a) Example of a 4-slice arc-shaped shim magnet. (b) Example of a 2-slice arc-shaped shim magnet. (c) Connection to the original magnet framework. (d) The arc-shaped structure is a symmetric partial Halbach array, defined by three parameters: arc radius, starting angle, and ending angle. (e) Each shim-magnet cube's location and orientation follow the Halbach distribution, with the orientation angle $\gamma_i$ being twice the location angle $\beta_i$.

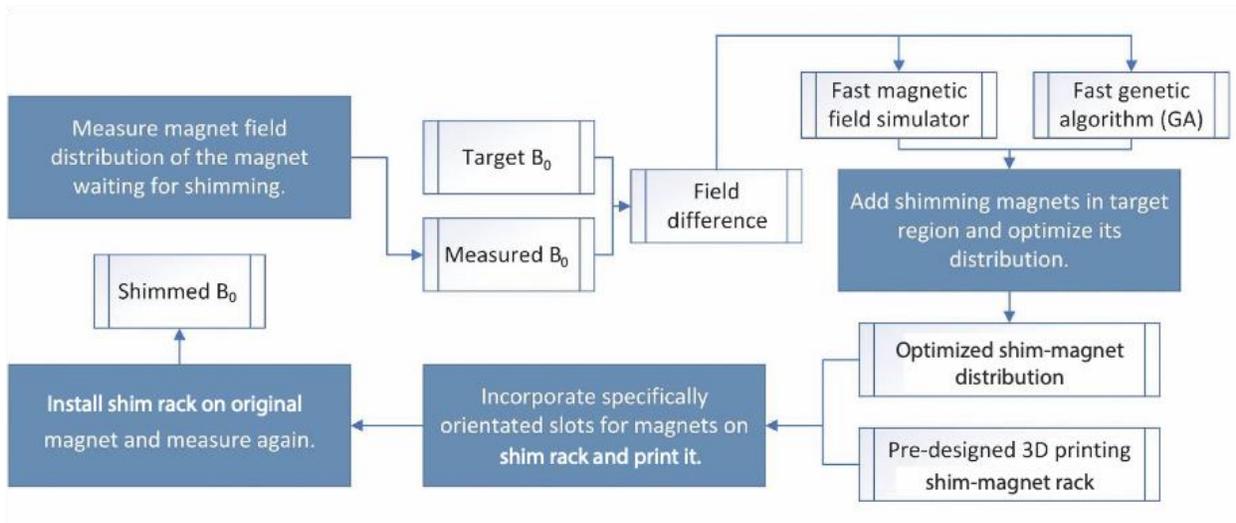

Figure 5. The entire workflow of the 3D printing integrated magnet shimming process.

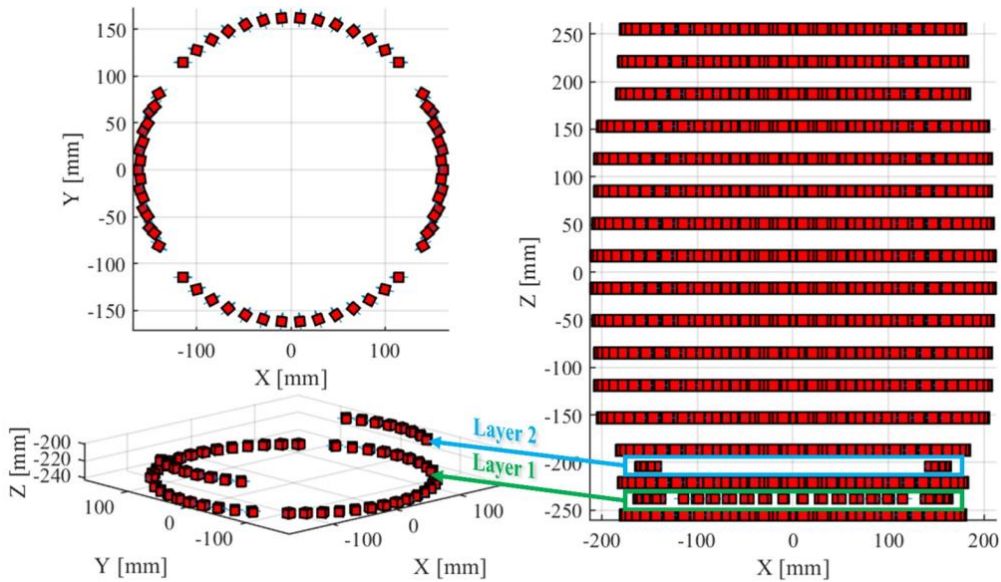

Figure 6. Optimized arc-shaped shim-magnet distribution. The design includes 6 arc regions with 2 layers, inserted into the slots between layers 1 and 2, and layers 2 and 3 of the original magnet, using 62 N52 magnets ($B_r$ = 1.42 T). Each magnet cube measures 10 × 10 × 10 mm³. The angular ranges for the first layer are: (-45°, 45°), (135°, 225°), (60°, 120°), and (240°, 300°). For the second layer, the angular ranges are (65°, 115°) and (245°, 295°). The radii of the first and second layers are 157 mm and 155 mm, respectively.

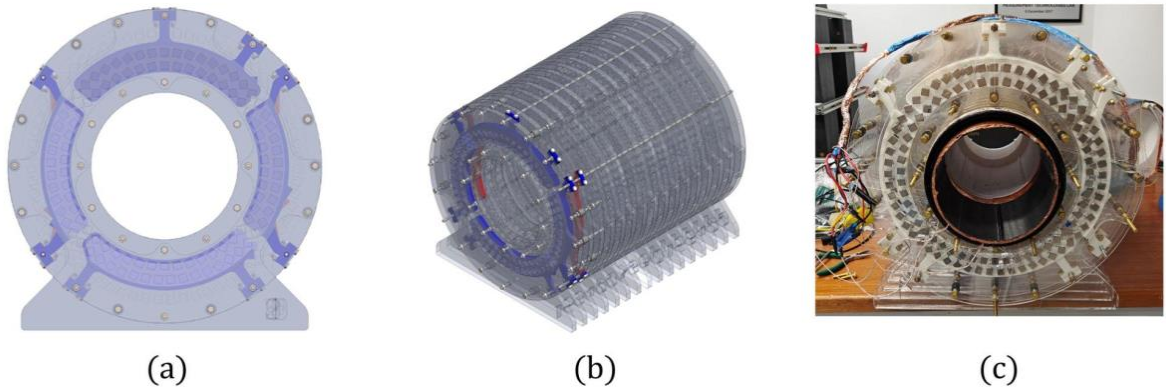

Figure 7. Integration of 3D-printed shim racks with the magnet housing. (a) Front view of the magnet housing showing the connections between the shim racks and the original magnet. (b) Isometric view of the magnet housing with the shim racks installed. In this study, 6 shim racks were added: 4 blue racks between layers 1 and 2 of the original magnet, and 2 red racks between layers 2 and 3. (c) The fully assembled sparse Halbach magnet array with the shim racks installed.

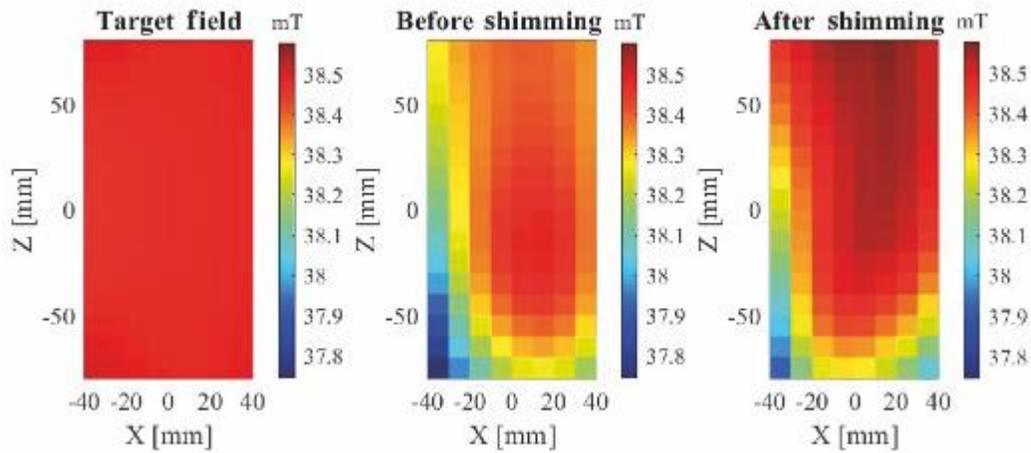

Figure 8. Impact of shimming on the magnetic field distribution. (a) Measured magnetic field distribution before shimming. It shows significant deviation from the target within the selected shimming region. (b) Target magnetic field distribution. (c) Measured magnetic field distribution after shimming. It demonstrates a closer match to the target and an increase in average magnetic field strength.

Table 4. Properties of the ezyMRI hackathon magnet.

| Property | Before optimization | After optimization |
|---|---|---|
| Total PM block layers | 16 | 18 |
| PM block number | 1904 | 1966 |
| Field strength (mT) | 38.32 | 38.42 |
| Homogeneity (ppm) | 194 | 171 |

## 4. Gradient coils

Resistive coils are used to generate spatially varying magnetic fields for spatio-temporal encoding of the NMR signal according to Eqn. (2). Linear magnetic fields used for spatial encoding are usually referred to as "gradients" instead of "gradient fields" $\mathbf{G}\ \mathbf{r}$, although the gradient coils generate magnetic fields that add to the magnetic field $\mathbf{B}_0$ of the MRI magnet (section 1.4). Magnetic field gradients are crucial for signal localization and constitute the main difference between an NMR spectrometer and an MRI device.

### 4.1 Design process

The main magnetic field $\mathbf{B}_0$ is much stronger than the gradient field. Hence, only the z-component $B_z$ of the gradient field $\mathbf{G}\ \mathbf{r}$ parallel to the main magnetic field needs to be considered for optimization of the MRI gradient coils as long as the gradient field and $\mathbf{B}_0$ point in the same direction [115]. The gradient coil was designed using the openly available CoilGen program [116,117] in MATLAB, which outputs a fully interconnected wire path as a 3D object. In CoilGen, the user defines the mesh parameters, e.g. geometry, number of divisions for the optimization surface, i.e. the representation of the cylindrical former on which coils will be wound, and the target volume. CoilGen produces an interconnected wire path from the level curves of the stream function on the optimization surface to optimize for gradient-field linearity in the target field. Users can set parameters of this optimization, such as the number of level curves used and regularization factor to find a balance between efficiency and field linearity suitable for the application at hand.

Wirepaths were output in the form of STL files, and the openly available software Blender [118] was employed to first generate a solid cylinder and then subtract the wire path to derive the coil former. Twelve Prusa i3 MK3S printers were operated in parallel to print sub-parts using polylactic acid (PLA) filaments from different suppliers. Formers were split into 3 to 6 subparts for two reasons: the 3D printers were not large enough to print the formers intact, and the speed afforded by parallelization was increased. The sub-parts were physically joined using a soldering iron through plastic welding. Once the formers were assembled, the coils were wound by simply affixing wire to the former grooves.

A gradient coil with three orthogonal channels along the spatial x, y, and z directions was designed, 3D printed, and assembled during the ezyMRI hackathon within less than 3 days.

Specifics of the design are elaborated in section 4.2, but overall, the design process was as follows:
- Specify the dimensions for each gradient layer according to the available space between the magnet assembly and the RF subsystem.
- Define mesh parameters of the target volume and optimization surfaces.
- Optimize regularization parameters of the three channels to derive wire layouts that generate accurate and equally efficient coil designs.
- Generate realizable 3D objects with grooves for winding coils with Litz wire.
- Segment formers to adapt to the available volume of the 3D printers.
- 3D print coil former parts.
- Fuse 3D-printed parts together using a soldering iron to get three individual coil formers.
- Wind each coil using a single Litz wire. Nevertheless, due to the wire available, soldered interconnections within the coil had to be made. The wire was fixed on the former with glue.
- Characterize each channel using a motorized Hall-probe setup.
- Combine the three channels and position them inside the magnet.

**4.2 Final designs**

Based on space availability, the x, y, and *z* gradient coils were designed with diameters of 204 mm, 209 mm, and 217 mm, respectively. A cylindrical mesh of the appropriate diameter, having 50 circular and 50 longitudinal divisions, was defined as the optimization surface for each coil. The target volume was a sphere of 70 mm radius with a linear gradient field in the desired coil direction. From the given optimization surface and target volume, coils were designed with wirepaths on level curves of the stream function. As the inversion of the Biot-Savart equation is an ill-posed inverse problem [119], Tikhonov regularization was used to find coil designs for all three coils that were practicable, accurate, and had roughly equal efficiencies. Regularization parameters were found using a trial-and-error approach with input values between 1000 and 10,000 and comparing efficiencies and the suitability of the wirepath for practical realization. Coil efficiencies of 0.8 - 0.9 mT/m/mA were targeted in the final design. This target represents the highest level of simulated accuracy of the least efficient coil, in this case, the *z*-gradient coil, and the other coils were designed to match this efficiency for ease of integration.

   3D wire paths from CoilGen were exported as STL files and subtracted from a cylinder on Blender [118] to create a cylindrical former with grooves corresponding to the correct wire paths. 3D models of formers for all three coils were sliced into 3 to 6 subsections (Fig. 9), 3D printed in parallel, and soldered together. Coils were wound by placing 2 mm Litz wire in the grooves of the formers and securing the wires with hot glue (Fig. 10).

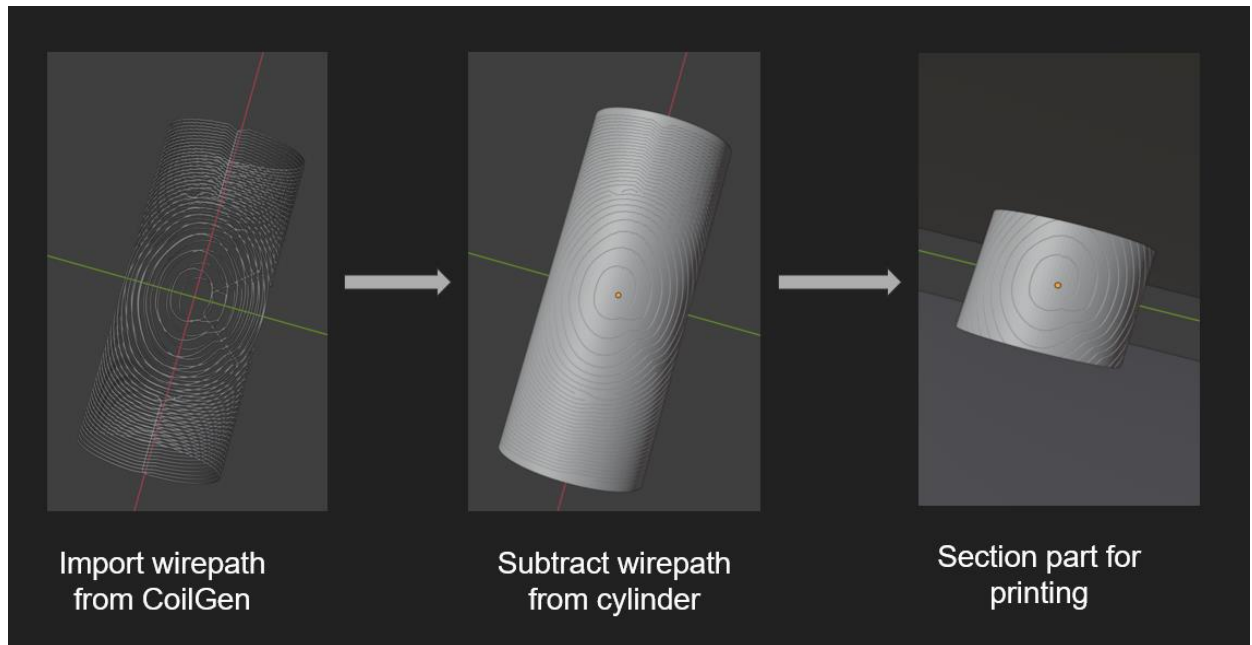

Figure 9. Blender workflow to create 3D printable former models.

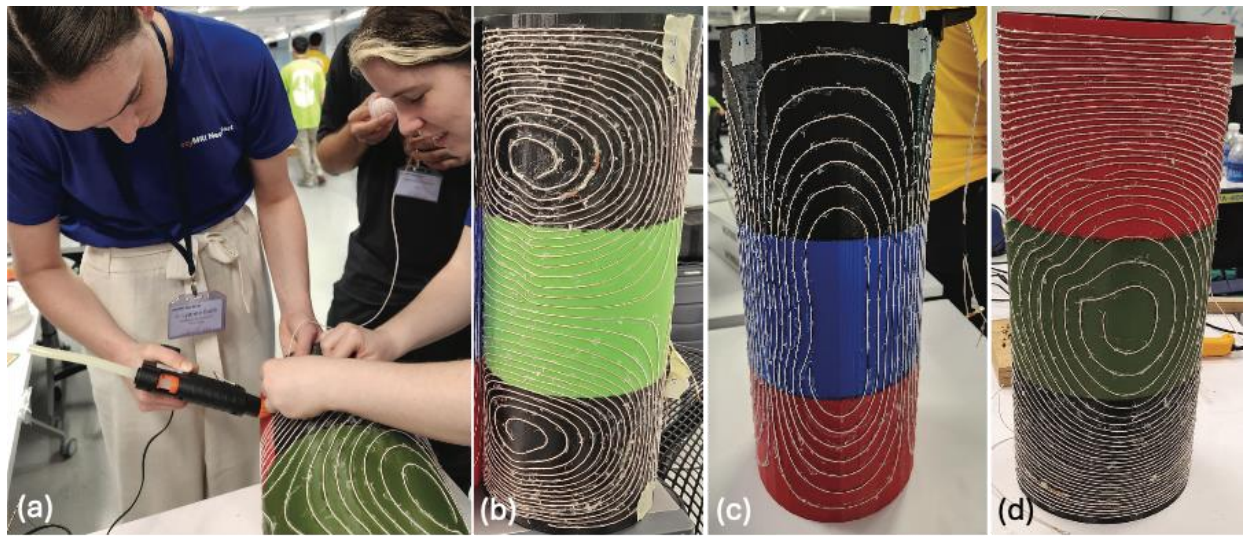

Figure 10. Litz wire was affixed to the cylindrical formers using hot glue (a) to wind the *x*, *y*, and *z* gradient coils (b-d).

To measure the magnetic field generated by the coils, a continuous 3 A current was run through each coil separately, and the resulting gradient field was sampled across the FOV by a motorized Hall-probe (Fig. 11). The resulting efficiencies of the *x*, *y*, and *z* gradients were 0.817 mT/m/A, 0.882 mT/m/A, and 0.832 mT/m/A, respectively. The coils' resistances/inductances were measured with an LCR meter, and their values were 1.29 Ω / 161 µH, 1.21 Ω / 72 µH, and 1.27 Ω / 250 µH, respectively (Table 5). After characterization, the three coils were fit together concentrically without additional physical joiners due to the tight fit between coils. The tight fit also alleviated the need for additional connector joints between the gradient coils and the main magnet.

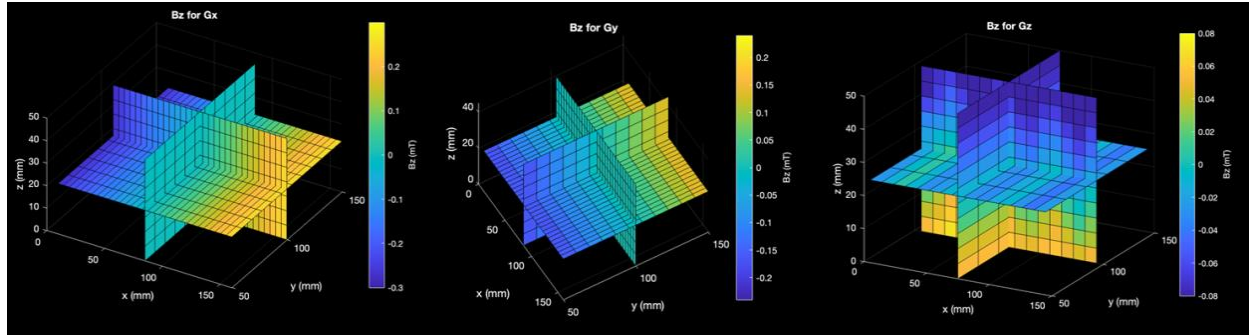

Figure 11. Magnetic field measurements of the gradient coils on the central planes in the FOV.

Table 5. Measured gradient-coil parameters.

|  | Efficiency (mT/m/A) | Resistance (Ω) | Inductance (µH) |
| --- | --- | --- | --- |
| $G_x$ | 0.817 | 1.29 | 161 |
| $G_y$ | 0.882 | 1.21 | 72 |
| $G_z$ | 0.832 | 1.27 | 250 |

Several avoidable complications arose that could be relevant for future designers. Originally, the *y* and *z* gradient coils were printed on a flexible printed circuit board (PCB) for simplicity of fabrication. These coils had a thickness of 0.2 mm, a copper layer thickness of 2 oz (0.035 mm), and dimensions of 172.4 mm × 522 mm for the *y* gradient coil and 150.17 mm × 522 mm for the *z* gradient coil. However, the resistance was ultimately too high and caused temperature concerns. An additional *x* gradient coil was wound with copper wire, instead of Litz wire, to compare the manufacturing methods. As expected, the copper wire was much less pliable than the Litz wire and thus fit less tightly into the grooves, which took longer to wind.

Due to time constraints, the regularization parameter for the *z* gradient coil was not fine tuned, and the final wirepath had zig-zags that complicated coil winding. The zig-zag wire path can be optimized by adjusting the *'normal_shift_smooth_factors'*, *'Tikhonov_factor'*, *'num_circular_divisions'*, *'num_longitudinal_divisions'*, and *'smooth_factor'* parameters in CoilGen through a trial-and-error approach. The efficiency, however, was as expected, and this caused no notable problems other than fabrication difficulty. Structurally, the gradient system was designed with a 0.5 mm clearance between concentric coils. In theory, this clearance could be sufficient if all wire was laid perfectly at the base of the grooves, but in practice, this led to difficulty in coil assembly as the concentric fits were too close.

## 5. Console

The console (aka control system, control electronics, or spectrometer) takes care of various critical functionalities in an MRI scanner. These are
- Time-synchronous execution of sequences of electromagnetic pulses (i.e. RF and gradient pulses, transistor-transistor logic (TTL) signals, etc.) to manipulate and encode the sample spins and spatial information.
- Acquisition and digitization of MR signals.
- Pulse-sequence development environment and compiler.
- Provision of a software environment for the user to interface with the machine.

To this end, an MR console typically encompasses (Fig. 2)
- an FPGA module,
- RF electronics (TxRx switch, low noise amplifier (LNA),
- tuning and matching elements, etc.),
- gradient control electronics (digital-to-analog converter (DAC) board,
- optional filtering and voltage amplification stages),
- a control computer,
- optional electronics for the control and monitoring of hardware or ambient variables (e.g. safety interlocks, coil temperature, fluid levels, etc.).

Whether high-power electronics are considered to be part of the console or not is situation-dependent and subject to interpretation. However, regardless of nomenclature, the RF and gradient power amplifiers have an influence on the console specifications and need to be considered during set-up. For instance, the TxRx switch must be compatible with the RF power amplifier (RFPA), gain and output, and the low-voltage DAC outputs for the gradient waveforms need to be made compatible with the GPA inputs, which are typically either single-ended or differential signals.

The console was built around the MAgnetic Resonance COntrol System (MaRCoS), which is an open-source MRI console running on the red Pitaya 122-16 SDRlab development board. Originally inspired by OCRA MRI [79], MaRCoS is designed to be easily programmable and possess no hard limitations on the sequence length or complexity. More details on the development and inner workings of MaRCoS are available in the literature [120] and on YouTube [121].

For the ezyMRI hackathon, the console team worked on five related sub-projects (Fig. 12). In the remainder of this section, the work conducted on all five subprojects is described, covering each phase up to the completion of the control system and the RF and gradient electronics. These tasks were carried out following the diagram in Fig. 13, where the interdependencies among tasks are indicated and the times at which the tasks were accomplished, as a tentative guideline for readers building or interested in building similar systems.

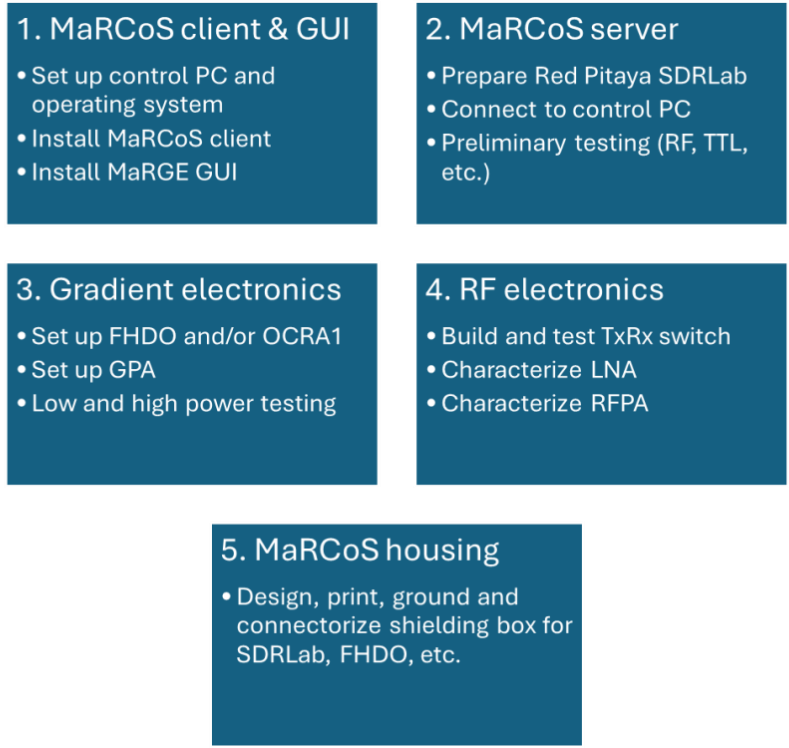

Figure 12. Main tasks.

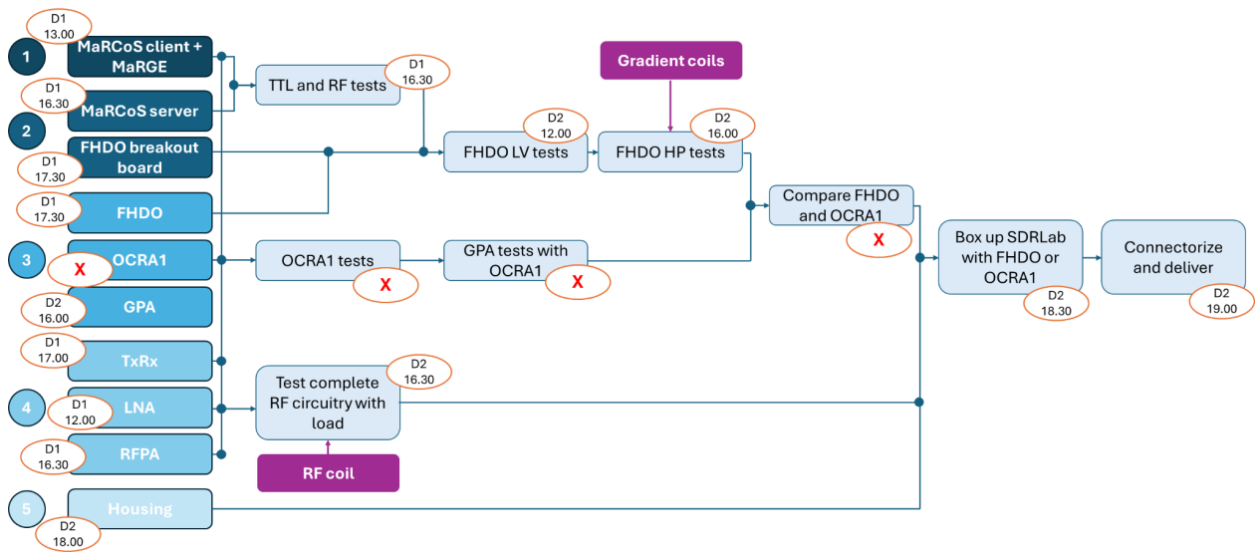

Figure 13. Flow chart of the tasks performed to finalize the console and prepare the electronics. The color-coded numbers at the left indicate the main tasks listed in Fig. 12. The ovals next to each box indicate when the tasks were finalized and give a sense of approximate time requirements. Those with a red X were not finalized because the OCRA1 printed circuit board was faulty (see below). The purple boxes correspond to components produced externally for the console team but were required for testing and calibration purposes.

The MaRCoS board is directly connected to an RFPA and an LNA for the respective transmit and receive RF coil channels, and paired with a GPA-FHDO board that drives the gradient coils. The console server handled the low-level real-time signal generation and sampling. High-level control, including pulse sequence design, is controlled by the MaRGE graphical user interface [80], a Python-based GUI application, and can be configured to run on Windows, MacOS, or Linux.

There are several prerequisites for installation and setting up of the console, in terms of both software and hardware. The main requirement for the hardware is a Red Pitaya SDRlab 122-16 standard kit [122]. The software requires a working Linux environment, such as Ubuntu or Arch Linux. Users with Windows OS can download the Windows subsystem for Linux or set up a virtual box. Secondly, it is useful to be familiar with using the command line interface to avoid errors while navigating the server system. Finally, the installation of Python and a number of libraries is necessary. Both MaRCoS and MaRGE have their own sets of libraries, listed in the `requirements.txt` text file for MaRGE. Prior knowledge of Python is required for users who wish to make changes to the system.

**5.1. MaRCoS client and MaRGE**

This subsection describes the setup process for the control PC, the installation of the MaRCoS client in the control PC, and the installation of MaRGE. In the hackathon an MSI 16 laptop running Ubuntu OS was used to control the MRI scanner. However, MaRCoS and MaRGE were also tested on various laptops running other Linux distributions, such as Windows and MacOS.

To set up the MaRCoS client, all dependencies were downloaded from the official MaRCoS GitHub [123] repository onto the client computer, and the `local_config.py` was configured on the marcos_client repository. For MaRGE, the installation instructions available on the MaRGE GitHub page were followed [124]. This involved installing Python and all required libraries as specified in the `requirements.txt` file in the MaRGE main directory and configuring the `hw_config.py` file according to the scanner's specifications. The client computer was fully set up by 13:30 hours on the first day (Fig. 13).

With both the client and the server ready (see Sec. 5.2), users could run MaRGE on the client computer (Fig. 14a) and execute a noise sequence even without attaching further hardware (Fig. 14b). This confirmed that MaRGE was now able to manage MaRCoS and run experiments. Then an RF pulse was applied and the output measured on the oscilloscope (Fig. 15). The MaRCoS client and the MaRCoS server were ready by 16:30 hours on the first day (Fig. 13).

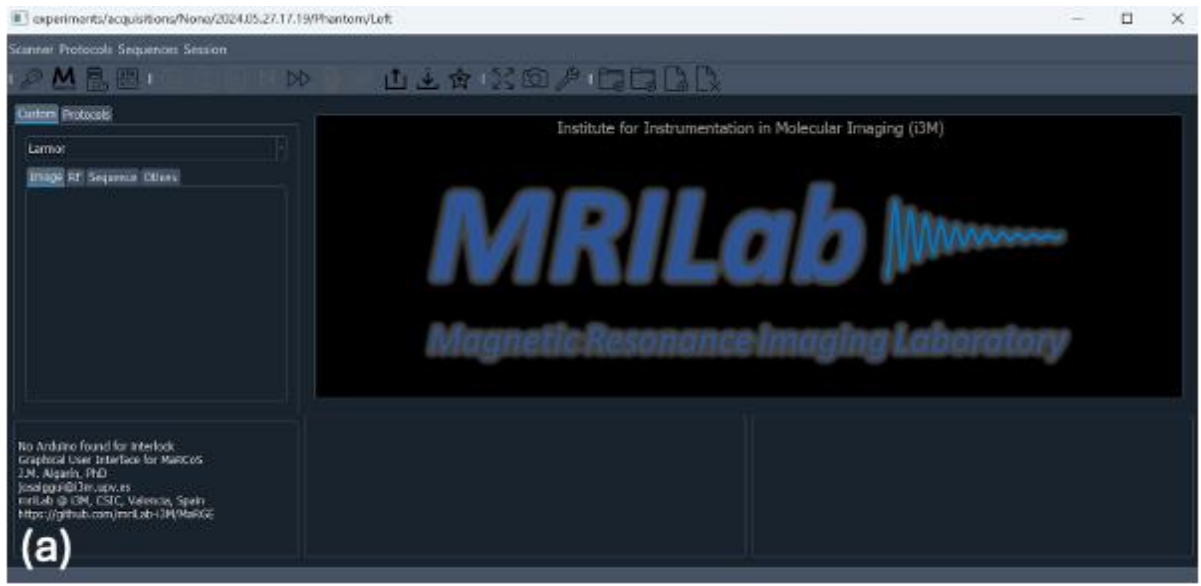

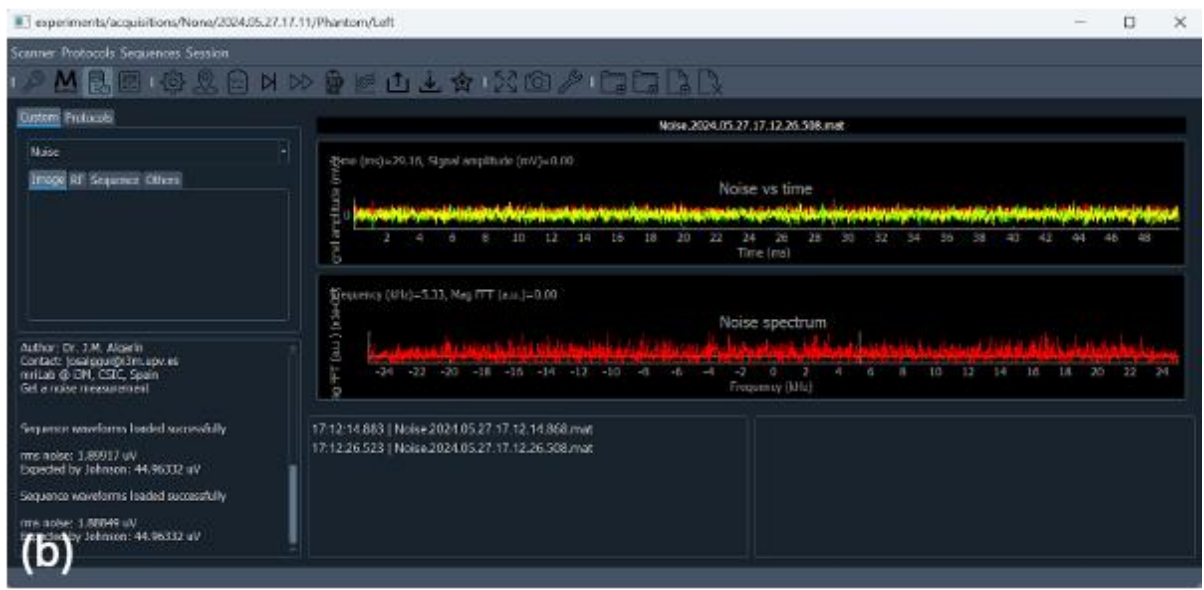

Figure 14. Screenshots of the MaRGE GUI. (a) Main window after launch. (b) Background noise check.

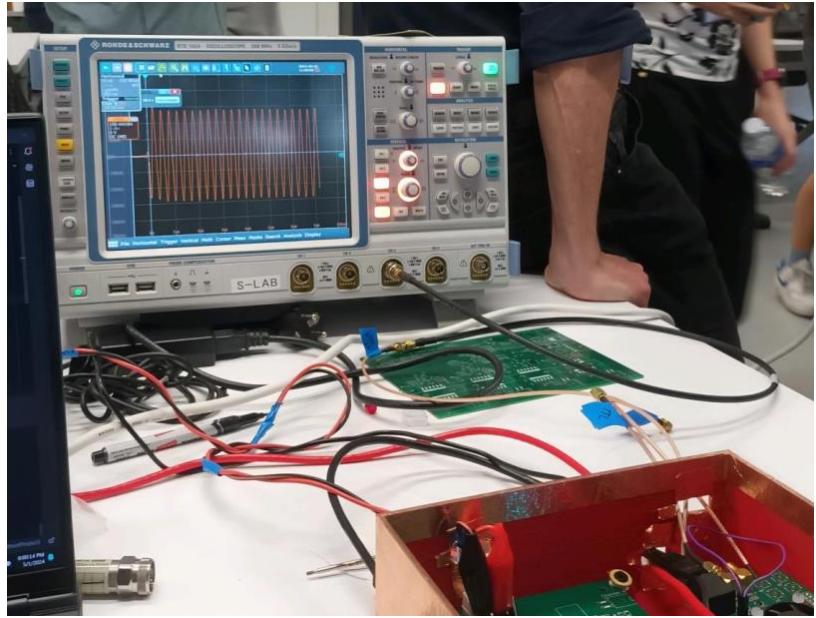

Figure 15. Picture of the RF pulse from the Red Pitaya using MaRCoS.

**5.2. MaRCoS server**

This subsection describes the process of setting up the MaRCoS server and establishing communication with a client computer. To install and set up the MaRCoS server, the official online documentation available on the MaRCoS wiki was followed [125]. During the event, a Red Pitaya SDRLab 122-16 standard kit was employed [126]. This includes the Red Pitaya board, an ethernet cable, and a secure digital (SD) card with a reader. First, a pre-built Yocto image was flashed on the SD card for use in the Red Pitaya. This process was done on a different computer from the one used to control the scanner, which also ran Ubuntu. Once the SD card was ready, it was inserted into the Red Pitaya, and the Red Pitaya was connected to the client computer via the ethernet interface to initiate communication.

To enable communication between the server and the client, the Red Pitaya OS required an IP address from a dynamic host configuration protocol (DHCP) server on the network. This posed a problem when connecting the board directly to the client computer via Ethernet, particularly on Windows and macOS. To resolve this, both the Red Pitaya OS and the Ethernet interface of the client computer were configured with a static IP address, ensuring smooth communication between the server and the client. To configure the IP address of the Red Pitaya, the interfaces file in '\etc\network\interfaces' of the SD card was modified following the indications on MaRGE wiki [127]. For the client computer, when using Ubuntu, the IP address was set by typing `ifconfig eno1 desired_ip_address` in the terminal as a root user. While using `ifconfig` may be straightforward, it required fixing the client computer's IP each time the computer was restarted. Fixing the IP address on the client computer permanently is OS dependent, and instructions for Ubuntu and Windows can be found in the MaRGE wiki. Once the IPs of the server and the client were properly configured, an SSH connection between the client and the server was possible by using `ssh root@ip_address` in the terminal.

Once proper communication between the client and the server was established, the MaRCoS repositories were downloaded onto the client computer used to configure the server. Then, the `local_config.py` file from the MaRCoS client folder was modified, where the IP address of the Red Pitaya was included. Finally, the `marcos_setup.sh` script was executed as indicated in the MaRCoS wiki [125]. The official MaRCoS image and bitstream files were found to be buggy and incompatible with MaRGE, so an earlier version of the bitstream files available in the MaRCoS extras [128] repository was used, specifically commit 433936c. This bug was later fixed and the last commit on the master branch of MaRCoS ran properly. After successfully executing the `marcos_setup.sh` script, the Red Pitaya displayed a blue LED, indicating that the bitstream between the server and the FPGA was active. The MaRCoS server was ready by 16:30 on the first day (Fig. 13).

### 5.3. Gradient electronics

Pulsed magnetic gradient fields encode the spatial information of the sample as required to reconstruct images. These gradient fields are generated by time-dependent electrical currents flowing through the gradient coils.

The Red Pitaya SDRLab board does not include sufficient suitable outputs to provide the three independent low-voltage signals that must be later amplified and converted into the high currents required by the pulse sequences. Instead, MaRCoS currently supports two different systems for gradient control, GPA-FHDO [129] and OCRA1 [130], connected through a serial-peripheral interface (SPI) bus. Both are reliable, low-cost (< 400 USD), open-sourced solutions.

The OCRA1 is a gradient board with ±10 V single-ended and ±20 V differential voltage outputs on four channels. It uses four 18-bit DACs running at their maximum SPI clock frequency of 35 MHz. Unlike the GPA-FHDO, the OCRA1 does not have a high-power stage and, hence, requires additional gradient power amplifiers, which users can choose according to their needs. Moreover, OCRA1 also includes an onboard Tx gate buffer for pregating the RFPA.

The GPA-FHDO is a gradient DAC incorporating a linear power stage that can supply ±10 A directly to the gradient loads. This board includes 4-channel 16-bit DACs, and the maximum SPI clock is around 40 MHz, which is limited by the SPI isolator. An adapter PCB for the Red Pitaya connects to the GPA-FHDO via Ethernet. It includes a buffer for the Tx-gate signal and a connector for an optional fan. A plugin module is also available for the GPA-FHDO, enabling the generation of a ±12 V bipolar output which allows use of the external analog gradient amplifiers.

For the hackathon, the initial idea was to construct and characterize both boards and compare their performance. To this end, it was intended to employ partially pre-assembled units, however both were faulty. In the FHDO, a NAND logic gate was found in port U20 instead of a NOR logic gate. This is trivial to fix under normal conditions, but these integrated circuits were unavailable during the event. Regarding OCRA1, there were clear indications of a short-circuit, however it could not be located and repaired. A different pre-assembled functional FHDO board was used but did not include the high-current amplifier stages, and had to be combined with the bipolar break-out board.

Once assembled and connected to the Red Pitaya, trapezoidal pulses were generated from MaRGE, and the low-voltage outputs were measured directly with an oscilloscope. By sweeping their amplitude in the unitless control in the GUI, the dimensionless numbers were calibrated to the real voltages provided by the board. As expected, all three DACs were highly linear, and the calibration factor corresponded to around 16 V/a.u. A home-made GPA was characterized by feeding the input channels with the calibrated voltages from the FHDO and connecting the outputs to a gradient coil. DC signals were sent from the DAC, and the current was measured with an Ampere meter.

**5.4. RF electronics**

*RF Power Amplifier (RFPA)*
In order to create a $B_1$ field that resonantly tips the magnetization, an RF power amplifier is needed to convert the low-voltage signal generated by the RedPitaya into high-power pulses driving the RF coil. To this end, the dimensionless RF amplitude control in MaRGE was swept, and the analog output was measured directly at the Red Pitaya with a 50-Ohm-terminated oscilloscope. A linear fit yielded a voltage amplitude of 228 mV/a.u. Secondly, the gain of RFPA was characterized with a Vector Network Analyzer (VNA), connecting port 1 to the RFPA input and port 2 to the RFPA output, all while pregating with a low duty cycle TTL signal. The center frequency of the VNA was set to 1.5 MHz, and it was made sure that the input signal never exceeded 0 dBm. An S21 measurement directly characterizes the RFPA gain, which can be calculated as 20·log(S21) in dB.

*Low-Noise Amplifier (LNA)*
The signal induced by the precessing spins in the RF antenna typically ranges from -70 to -50 dBm, and an LNA can amplify this signal to maximize the ADC's dynamic range (250 mV discretized in 16 bits for the Red Pitaya). A Narda-Miteq LNA was characterized by generating a 1.5 MHz sinusoidal waveform with a function generator. 50 Ω attenuators were then inserted in the RF chain to bring the amplitude close to -50 dBm. An LNA gain of 27.1 dB at 1.5 MHz was measured, slightly higher than the expected 26 dB.

*TxRx switch*
A single RF coil was used for both $B_1$ excitation and signal detection, which required a switching circuit to prevent powerful RF pulses from damaging sensitive readout electronics during transmission and ensured that the full signal reached the ADC during reception. For the ezyMRI hackathon, the integration team assembled a homemade, passive TxRx switch based on a λ/4 lumped-element circuit. Specifically, this was realized with an inductor and two capacitors, arranged into a π network [131] and resonating at 1.5 MHz.

## 5.5. MaRCoS housing

Housing for the console hardware contained the SDRlab 122-16 board and the GPA-FHDO to shield the console from external electromagnetic interference while keeping the entire console system compact. The main box with five sides (Fig. 16) was designed and printed along with a lid. Holes were drilled to accommodate all required connectors (RF, Tx, and Rx, ethernet communication with the control PC, low-voltage gradient lines, and power from an external benchtop power supply). Finally, the outer surfaces were wrapped in conductive copper tape and electrically connected to the grounding shield between the RF coil and the gradient coils.

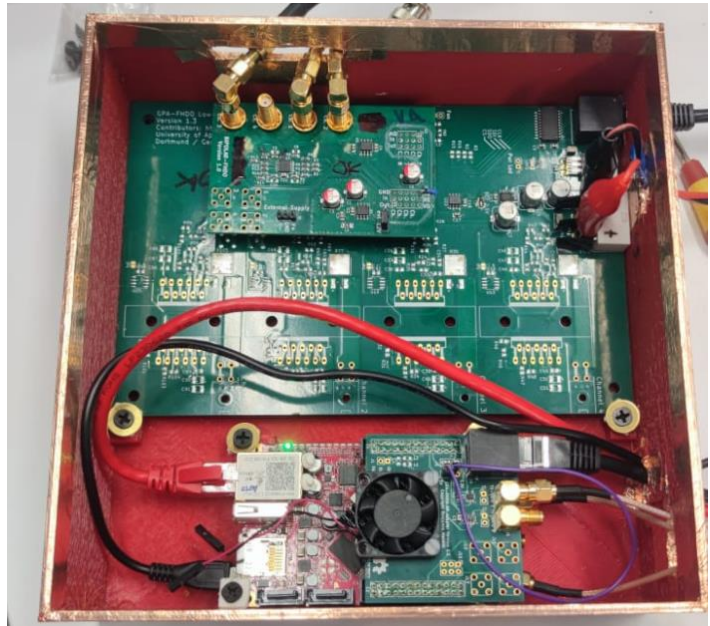

Figure 16. Photograph of the MaRCoS housing built during the hackathon. The Red Pitaya is at the bottom and the FHDO and bipolar break-out boards on top.

## 6. RF coil

### 6.1 RF coil - background and criterion

An extensive discussion of coil geometry, the effects of the conductor used in coil construction, and the effects of RF shielding can be found in the literature [132]. Briefly, the transverse orientation of the $B_0$ field in a Halbach-based magnet suggests that the geometry for the RF coil preferably is a solenoid which produces $B_1$ field coaxial with the cylinder axis of the magnet. The coil can be cylindrical, elliptical or suitably shaped to conform with the geometry of the human head [90]. As shown by Hoult [133], when the length and diameter of the coil are similar, the solenoid is significantly more sensitive than other RF coils, such as a saddle-shaped coil or a birdcage resonator, which produce less homogeneous $B_1$ fields in the transverse plane. Often a variable winding pitch is used to increase the homogeneity of the solenoid, with a higher wire density at the ends of the coil [134]. For the hackathon, the solenoid configuration has been used in both transmit and receive modes.

To avoid a reduction in coil sensitivity caused by phase changes in the $B_1$ field, the axial length of the solenoid coil should be much smaller than the wavelength. In addition, dielectric losses occur in the sample if an excessive number of turns is used in the solenoid coil. Capacitive segmentation can reduce both effects and is recommended for use even at very low fields.

Coils can be constructed from thin copper tape, circular copper wire, hollow copper tube, and Litz wire. From a theoretical point of view, the optimal choice in terms of reducing losses in the conductor would be a copper tube with a diameter of a few centimeters, or Litz wire with several thousand individual filaments. With appropriate capacitive segmentation, this would give a coil with an unloaded quality factor $Q$ of several hundred, corresponding to an effective RF coil bandwidth of ~10 kHz. Given the practical challenges of constructing RF coils with different dimensions at the hackathon, along with the known fact that the magnet homogeneity will be tens of thousands of ppm without shimming, and the field likely be drifting with changes in temperature, it was decided to use copper tape and thin copper wire for the construction of the coils. This resulted in $Q$ values closer to 100 and effective bandwidths of several tens of kHz.

An RF shield consisting of a thin solid copper sheet is required to reduce the interaction between the RF coil with its external environment and avoid extra noise in the detected signal [135]. The relatively small diameter of the magnet means that the RF coil is very close to the inner Faraday shield. Image currents are induced outside of the shield, effectively introducing a negative mutual inductance into the RF coil circuit and increasing the resonance frequency [135]. Therefore, the RF shield needs to be in position for final coil tuning and impedance matching. To this end, an L-network circuit with two variable capacitors was used. The test sample consisted of three 600 ml bottles filled with ~1% copper sulfate doped tap water (Fig. 17b), which did not exert a heavy load on the coil and rendered retuning of the sample unnecessary.

**6.2 RF coil - specifications, constructions, and testing**

A low-field RF transceive coil resonating at 1.618 MHz was constructed, which consisted of 8 turns of copper wire wound around a plastic cylinder with 145 mm inner diameter (Fig. 17a). The spacing between neighboring turns of the wire was empirically set to be uniform and approximately 10 to 15 mm apart. A PCB circuit board with carefully selected electronic components was soldered to both ends of the copper wire for coil matching and tuning. This solenoid coil was designed for an 80 mm diameter of spherical volume (DSV) (Table 1). For simplicity, the PCB board directly connected to the RF coil was finally used for tuning and matching only, as the control electronics for the TxRx switch was implemented by the console team.

To ensure the RF coil was precisely tuned to the targeted resonance frequency and matched to 50 ohms for optimal power transmission, the resonance circuit was iteratively resoldered to adjust its resonance characteristics measured with a network analyzer (Fig. 17b). The iterative process continued until the resonance frequency aligned with the Larmor proton frequency that matched the field of the constructed magnet, achieving approximately -20 dB in the S11 parameter. For these tests, three 600 mL bottles of drinking water were used as the load phantom, and a short-axis RF shield was temporarily placed around the RF coil.

Once properly tuned and matched on the bench, the RF coil was inserted into the constructed magnet (Fig. 18a,b), and FID signals were observed without gradients using the integrated electronics platform. The S11 spectrum of the RF coil inside the magnet closely resembled the bench measurement, and FID signals were observed on the user interface. Subsequently, the plastic support of the RF coil was manually polished (Fig. 18c), and the RF coil was permanently fitted into the long-axis RF shield (Fig. 17c). Eventually, the RF coil was combined with the other subsystems that constituted the low-field MR imaging instrument. Some tools helpful for making coils and reducing their noise are a short-axis RF shield for noise reduction with cut copper sheets and capacitors soldered to reduce eddy currents, a long-axis RF shield for improved coverage and stronger noise reduction, a balun for testing the RF noise reduction, and a sniffer coil for RF circuit debugging (Fig. 19).

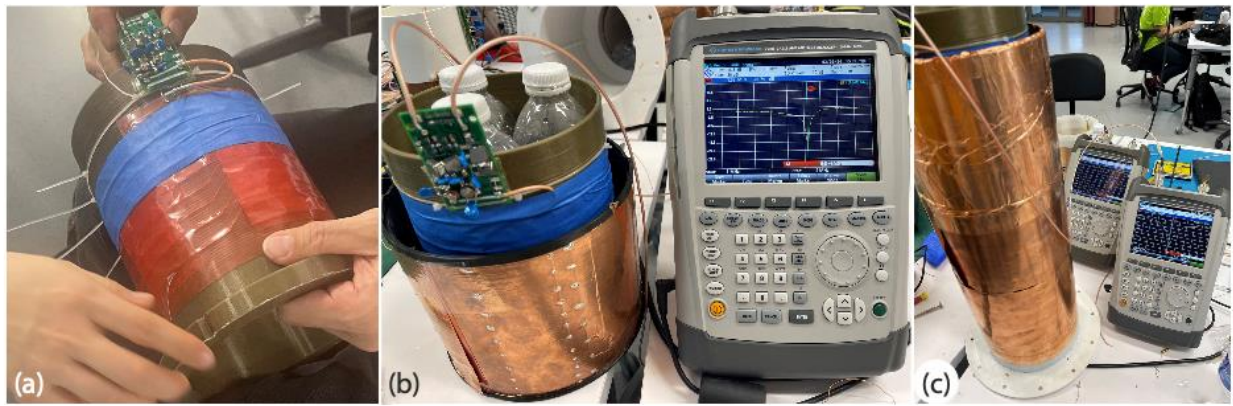

Figure 17. RF coil during construction. (a) Solenoid RF coil with a matching and tuning circuit board. (b) The constructed RF coil was tested with a network analyzer to examine whether the resonance frequency of the coil was properly tuned to the one determined by the magnet, and if its impedance was matched. A short-axis RF shield was used during these tests. (c) The RF solenoid coil was eventually pushed into the long-axis RF shield for more thorough coverage as the final design and tested again with the network analyzer.

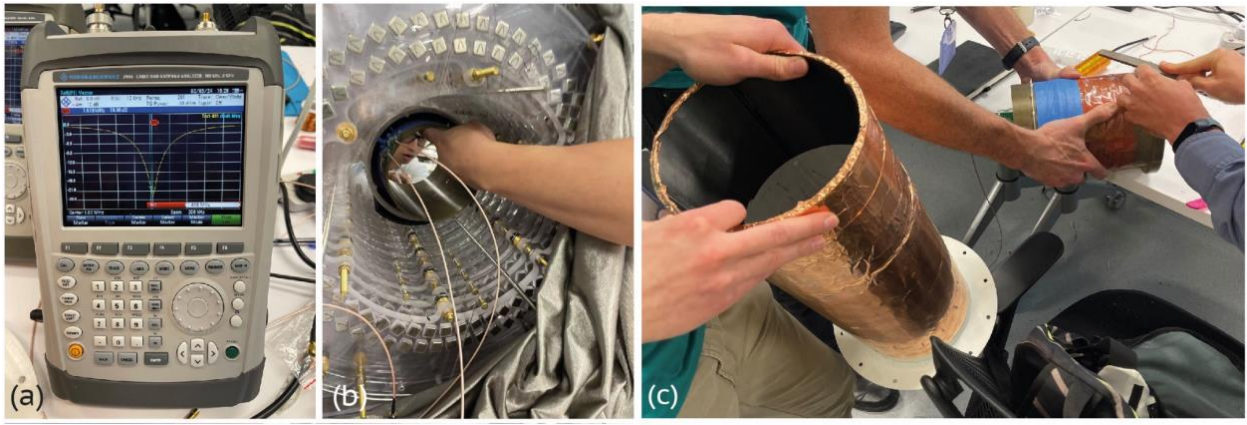

Figure 18. The RF coil during testing. (a) Network analyzer to examine tune and match properties of the RF coil. (b) The matched and tuned RF coil was placed into the magnet to examine the RF signal quality. (c) The matched and tuned RF coil was manually polished to fit into the long-axis RF shield.

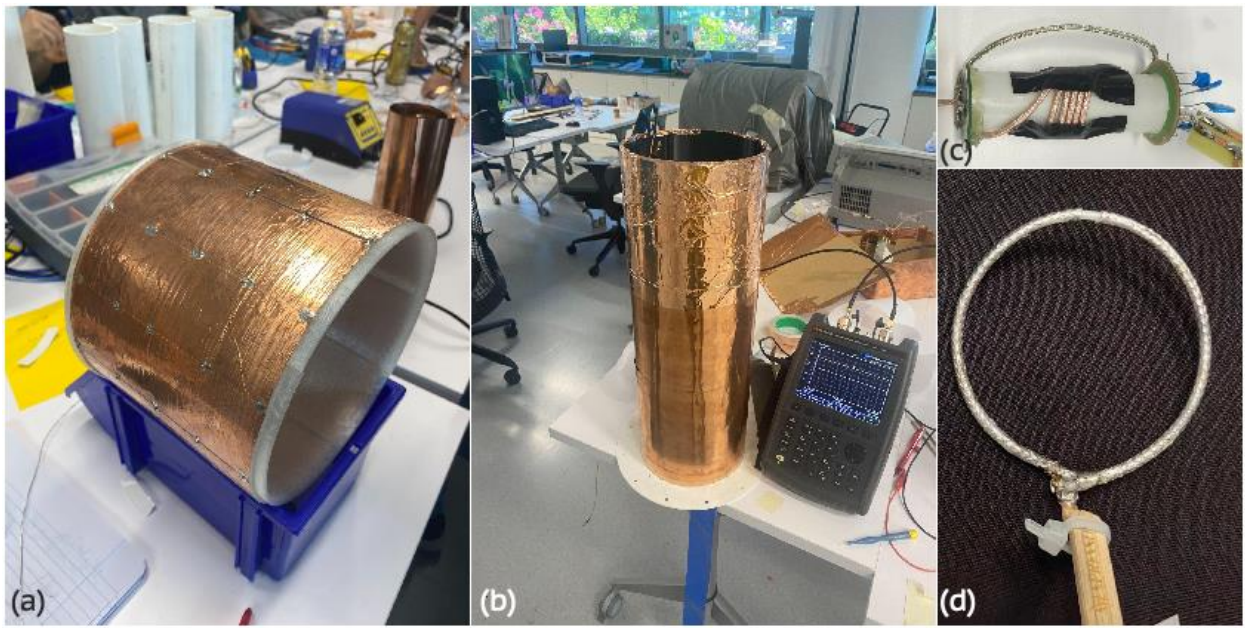

Figure 19. Tools made to facilitate the RF coil debugging, construction, and operation. (a) Short-axis RF shield for noise reduction at low-field with cut copper sheets and capacitors soldered to reduce eddy currents. (b) Long-axis RF shield for improved coverage and stronger noise reduction. (c) Balun for testing the RF noise reduction. (d) Sniffer coil for RF debugging.

## 7. Integration and testing – a critical analysis

All components that needed to be integrated to make up a low-field MRI instrument are visualized in Fig. 2. Details of the individual parts (Fig. 20) are described in the previous sections. The physical integration involved sliding the gradient set into the bore of the magnet, inserting the RF shield inside the gradient coils, and finally inserting the RF coil at the center of the magnet. Due to the transverse direction of the **$B_0$** field, care had to be taken to ensure correct orientation of the gradient coils with respect to the direction of the **$B_0$** field. The RF coil had been tuned and matched with the RF shield in place, leading to no measurable frequency shifts associated with the presence of the gradient coils and main magnet. A conductive shielding cloth was placed over the entire assembly as a rough method of reducing electromagnetic interference. This also acted as a grounding point to which the ground of the cable from the RF coil was attached. Further grounding points were established on the RF shield, and the loaded RF coil was re-tuned to the previous resonance frequency determined by the magnet. During integration, both the gradient chain (green blocks in Fig. 20) and the RF chain (blue blocks in Fig. 20) were tested.

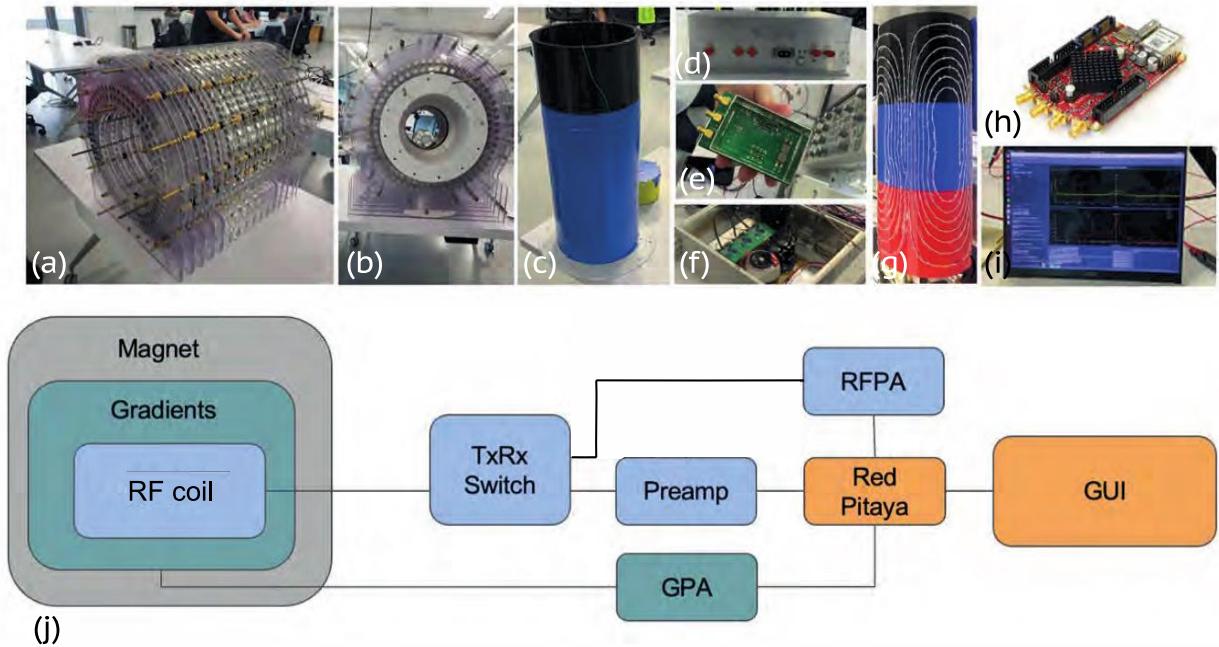

Figure 20. The components to be assembled. (a,b) Halbach-Array. The receive and transmit path consisted of the RF coil and RF shielding (c), an RF power amplifier (d), a passive TxRx switch (e), and a preamplifier. The gradient system consisted of the gradient amplifier (f) and gradient coils (g). The console system comprised (h) a Red Pitaya 122-16 SDRlab and (i) a laptop. (j) Diagram of the entire setup.

## 7.1 Gradient chain

The spectrometer, the gradient power amplifier (GPA), and the gradient coil are connected as shown in Fig. 20. The spectrometer provides specific gradient waveforms. The gradient chain was tested by monitoring the current waveform through the monitor port of the GPA. Distortions of waveforms were seen due to the mismatch between the GPA and the coil. They were mitigated by adjusting the matching of the GPA to the coil.

The low-voltage DAC outputs from the FHDO are single-ended (signal and ground), however, the GPA expects differential-mode inputs (positive, negative, and ground). In this case, the signal should be connected to the positive pin and the single-ended ground to the negative input. The grounds MUST NOT be connected to avoid interference with the feedback loop that stabilizes the GPA. During system integration, this mistake happened, and the gradients were driven by the GPA as soon as it was powered on, even when there was no signal at the input. Grounding the negative inputs caused the GPA to supply large currents to the coil, which were subsequently switched off by the interlock safety circuits. This process repeated at audible frequencies.

Another aspect to consider is the potential oscillation of the gradient amplifiers caused by the respective impedances of the output of the gradient amplifier, the input of the gradient coils, and the gain of the feedback loop used in many gradient amplifiers. The resistance and inductance of gradient coils for this type of application are typically ~0.5 Ω and 200 µH, respectively. Gradient amplifiers have specifications for the range of impedances that can be driven without the occurrence of oscillations. As it turned out, two of the gradient coils did oscillate, which could be heard by a high-pitched noise. For sophisticated amplifiers, it is possible to overcome this by altering the dynamics and gain of the feedback loop, or by changing the output impedance of the output matching network. However, these options were unavailable so resistance was added to the gradient coils to prevent oscillation.

**7.2 RF coil and amplifier**

The output impedance of most RF amplifiers is 50 Ω; therefore, for maximum power transfer the input impedance of the RF coil was also 50 Ω. However, at the low frequencies determined by the low-field magnet, there was actually no penalty for impedance mismatch if a high-impedance preamplifier was used in the receive chain. If a very high $Q$ coil is employed, for example, by using Litz wire instead of the conventional wire used in this case, the RF amplifier must be able to withstand significant reflected power in case the coil frequency drifts, or RF pulses are applied significantly off-resonance. Ideally, the RF amplifier can be gated on and off using TTL pulses from the spectrometer so that transmitter noise cannot enter the receive chain even when the amplifier is gated off.

**7.3 Receive chain sensitivity**

The receive chain was tested by coupling a resonant, sinusoidal signal inductively into the receive chain using a Keysight N9810A signal generator (Fig. 21). By incrementing the output power of the frequency generator in steps, the minimum power detectable by the receive chain could be determined. A signal with a power of -70 dBm could be measured, which was sufficient for detecting the voltage induced by the precessing magnetic dipole moments.

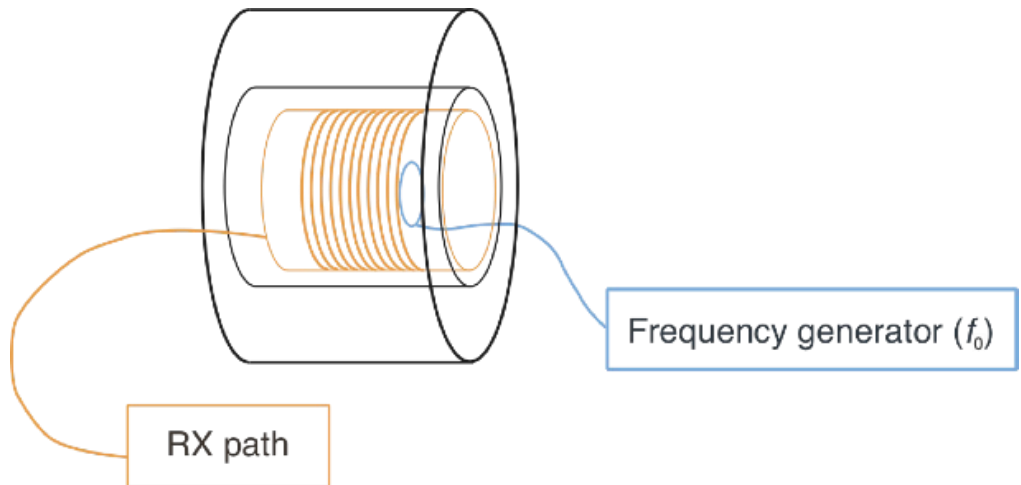
Figure 21. Evaluation of the receive chain by coupling a sinusoidal signal with the magnet's resonance frequency.

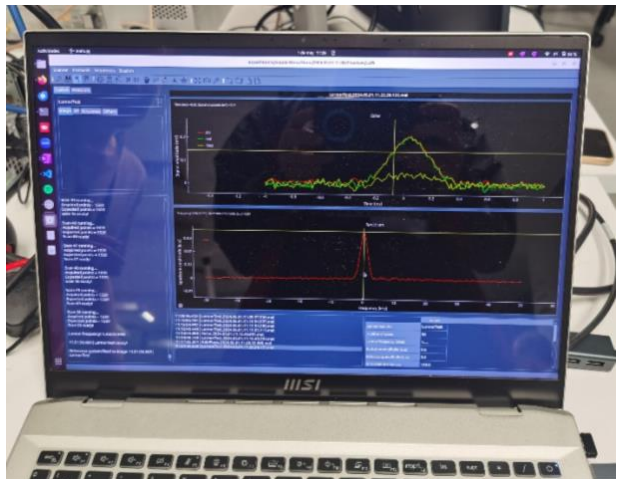
Figure 22. The echo signal acquired after the components were connected and the RF chain was tested.

The receiver noise was tested with the spectrometer using a 50 Ω load connected to the probe port as a "gold standard". Measurements were taken with RF coils and gradients connected, but the corresponding amplifiers were first turned off and then turned on. Ideally, the noise level would not change, but in practice, there was a large increase which indicated insufficient blocking of RF noise by the Tx/Rx switch, with the RF amplifier not TTL-blanked, and a lack of filters on the gradient feeds.

### 7.4 Calibration tests and sequences

Once all components were connected the RF chain was tested by executing a spin-echo sequence and acquiring an echo (Fig. 22). Pulse length and power as well as the flip-angle and the $B_1$-field uniformity across the sample were calibrated with 450°/90° tests. The background noise was checked with a simple pulse-acquire sequence without the RF turned on, while the input was terminated with a 50 Ω load. This step was subsequently repeated with all the components turned on.

**7.5 Results and potential improvements**

Although each component worked well individually, the SNR was significantly lower than expected when the entire system was turned on. This was attributed to different factors due to the given options. Although not an exhaustive list, in terms of the electronics, some contributors to the noise experienced included
- using an RF amplifier that has blanking capability,
- improving the insulation between the RX and TX channel of the passive TxRx switch,
- incorporating in-line filters between the gradient amplifier, which should ideally be linear rather than differential, and the gradient coils,
- employing a common grounding point or grounding bus which can also be connected directly to a common ground of the building to reduce the influence of environmental RF signals leaking into the system.

Finally, the effectiveness of actions intended to decrease the noise performance needs to be investigated by, for example, comparing the noise level to Johnson noise as offered by MaRCoS.

**8. Product design – envisioning the future of low-field MRI**

The product design team was tasked with conducting background research on portable low-field MRIs, collaborating with other teams to address physical dimensions and constraints of the MRI instrument, while designing the user interface. Initially, the team was asked to 'design an ideal, futuristic way for imaging internal structures of a body' to encourage out-of-the-box thinking without engineering constraints. Shifting focus to a more practical, portable MRI design, the team then focused on key design aspects such as ease of use, user comfort during the scan, and aesthetics. Table 6 outlines the two main functions for the design.

Given the increase in pet ownership after the covid pandemic [136-140], the team settled on developing portable MRI for pets to diagnose brain and spinal cord issues, where detailed images of soft tissues are needed (Fig. 23). The portability allows for at-home pet services, enabling pet owners to accompany their pets during scanning. Adapting the MRI instrument to accommodate animals provides valuable insights to adjust the design to different body part shapes, sizes, as well as consider the behavior of non-cooperative subjects.

The team followed the design innovation framework for product development [139]: Discover, Define, Develop, and Deliver, applying tailored methods for each stage. The process combined design and technology, beginning with conceptual ideation that balanced technical feasibility with visual appeal (Fig. 23a). The focus was to address unmet medical needs and transform innovative ideas into feasible designs (Fig. 23b,c). During the Discover phase, the team conducted background research using journey maps of users and personas to gather insights. Pets were used as an 'extreme user group' [140] to push boundaries of product design, given that humans are typical users of the MRI. In the Define phase, they created functional models, activity diagrams, and framed "How Might We" statements to shape the design direction. The Develop phase utilized C-sketching for ideation and rapid prototyping to iterate and refine ideas. Lastly, in the Deliver phase, CAD modeling, rendering, and 3D printing were used to create the final product (Fig. 23d,e). Each phase provided new insights and enabled the team to tailor the design to specific users, and the unique scenarios associated with their needs.

Table 6. Functions and criteria envisioned for the future of low-field MRI

| User friendliness | Feasibility |
|---|---|
| <ul><li>Comfortable and cushioned</li><li>Receive coils should be specially designed for different body parts and sizes</li><li>Scanner dimension should not be too long</li><li>Interface is operator-friendly</li><li>Adjustable angles for different positions</li><li>Portable instrument, allowing it to be brought to the user</li></ul> | <ul><li>Function under common voltage</li><li>No extra shield</li><li>Compact</li><li>Light weight</li><li>Low cost</li><li>Image quality sufficient for diagnostic purposes</li><li>Safety</li><li>Regulatory approvals</li><li>Smart</li></ul> |

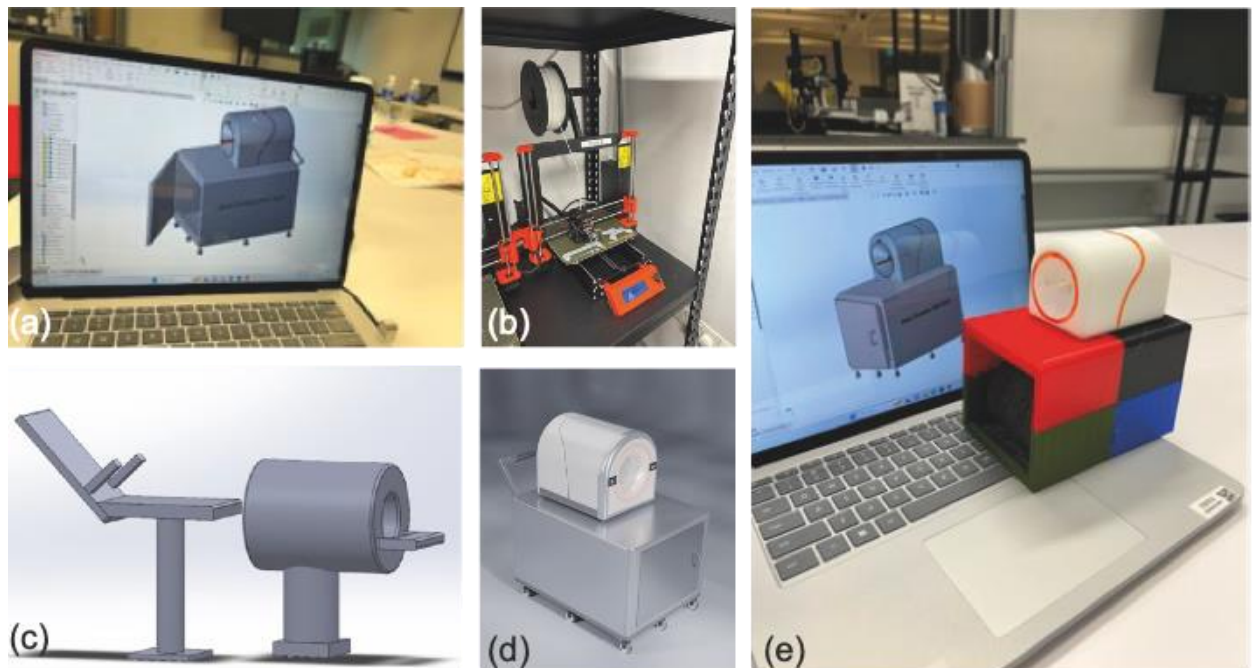

Figure 23. Designing of a portable at-home pet MRI instrument. (a) CAD modeling. (b) Prototype iterations through 3D printing. c) New ideas during modeling: a low field extremity scanner system. (d,e) Final design: mini model and rendered instrument.

Table 7. Design considerations

| Feature | Property |
|---|---|
| Aesthetics | Incorporation of animal-like features (biomorphism) to product exterior to evoke sense of familiarity and appeal to increase user attraction |
| Table | Designed for easy tool placement and maintenance |
| Handles & wheels | Optimized number of large wheels for mobility and enhanced stability. |
| Bed | Automated and adjustable, with a tiltable feature to enhance patient comfort and ease of access |
| Material/texture | Smooth finish with soft touch for comfort |

The team addressed these key questions through the design process (Table 7): How does one design for extremities and different body parts of humans and pets used in MRI? Given the weight of the low-field MRI, how should the interface move or tilt to optimize imaging of different body parts? How can the MRI machine ensure patient comfort, particularly when a stable position is required for prolonged scanning periods? How can the interface be made more user-friendly for clinicians or medical staff? What materials should be selected to avoid interference with the image acquisition, such as non-magnetic, lightweight, permeable, or biocompatible materials?

## 9. Summary

Low-field NMR is an area of activity where instrument development is still accessible to scientists and engineers at academic institutions outside profit-oriented commercial enterprises. Particularly attractive in this regard is diagnostic low-field MRI, because of its enormous potential benefits to human society and the many opportunities unfolding through innovative permanent magnet technology, 3D printing, affordable powerful personal computers, and software resources for acquiring and processing images at low signal-to-noise ratios in imperfect magnetic fields. The 2024 ezyMRI hackathon at Singapore University of Technology and Design was the second exercise of building a low-field MRI instrument from scratch in a short period of time by several teams of enthusiasts who came together from all parts of the world. Relying on simple hardware and open-source software, a functioning MRI machine was assembled in three days following a one-day training event with introductory lectures. At the last minute of the last day, a 2D projection image of two water bottles could be acquired, which demonstrated the essential functioning of the instrument albeit at low image quality. The different steps taken by the teams involved are reported in this work. Although the open-source documentation was a fundamental enabler for this project, it is not always easy to follow. Moreover, the ongoing evolution of computer technology along with updates of software environments necessitates permanent maintenance of the open-source documentation, creating a time-lag between hard- and software used and the available documentation, so that expert help will always be needed for the novice engaging in building a low-field MRI machine. The growing community interested in this subject will eventually remove more and more obstacles along this path. This report is intended to motivate and guide future such efforts.

**Acknowledgements**

Andrew Webb is funded by the ERC Advanced Grant PASMAR and Rui Tian is funded by the ERC Advanced Grant (No. 834940). The following teams and individuals collaborated in the ezyMRI hackathon: *Console team:* Mentor Joseba Alonso. Members Dan Xiao, Eddy Solomon, Gabriel Zihlmann, Huseyin Enes Candan, Irena Zivkovic, Jose Borreguero Morata, Jules Vliem, Li Kuo Tan, Steven Winata, Teresa Guallart Naval, Xinyu Ma, Yiju Guo, Yonatan Urman. *Design team* Mentor: Jacob Kang. Members: Can Xu, Qiongrou Ye, Ruiqi Jin, Sharada Balaji, Xiaohan Liu, Yiming Liu, Zhonghai Chi. *Gradient team:* Mentors Sebastian Littin, Maxim Zaitsev. Members Daniel Abraham, David Corcos, Jiahui Ding, Punith Bidarakka Venkategowda, Isabelle Zinghini, Lyanne Budé, Shah, Zachary, Yujun Ling, Yuwan Wang. *Integration team:* Mentors Wenwei Yu, Jie Liu, Zhiyong Zhang, Andrew Webb. Members Fabian Bschorr, Kay Igwe, Rongxing Zhang, Xinxin Li, Yueqi Qiu. *Magnet team:* Mentor Wenwei Yu. Members Bhairav Mehta, Chang Sun, Gonzalo Rodriguez, Jaladhar Neelavalli, Mark Nishimura, Naoto, Fujita, Neale Wiley, Pohchoo Seow, Shengyuan Wu, Shiwei Yang, Víctor Ramírez Olate, Tingou Liang. *RF team:* Mentors Joe Li, Paul Cassidy. Members Abel Worku Tessema, Anas Bachiri, Carlos Castillo Passi, Efrat Shimron, Hua Guo, Lachlan West, Minxuan Xu, Nayebare Maureen, Ronald Amodoi, Rui Tian, Ruian Qin, Sergey Korchak, Suen Chen, Tao Yun, Xiangzheng Kong, Yingyi Qi, Zhaoyang Hao, Ziye He, Ziyu Fu. *Event facilitating team*: Jing Han Heng, Junqi Yang, Anieyrudh R, Saumitra Kapoor, Christyasto Pambudi, Yifeng Jiang, Gan Bei Ru, Wong Jae Hann, Chia Pei Zhi, Tan Ying Yi, Stalin Thileepan, Wenjing Chu, Sok Yen Chong, Han Yat Siew, Eric Tan, Hilmi Yusoff. Support of the ezyMRI hackathon by Ngee Ann Kongsi (platinum sponsor), United Imaging and RFSGTech (silver sponsors,) and Resonint (bronze sponsor) as well as the support of food on the seminar day by the IEEE Singapore Sector are gratefully acknowledged.

**Author contributions**

Shaoying Huang: Organized the ezyMRI hackathon, coauthored sections 1.3 and 8, and edited the manuscript.
José Miguel Algarín: Developed MaRCoS and MaRGE, coauthored section 5
Joseba Alonso: Supervised the console team, coauthored section 2, and wrote section 5
Anieyrudh R: Coauthored section 3.2.1, handled 3D Printing, and assisted other fabrication efforts during the hackathon
José Borreguero: Participated in console and electronics assemblies, coauthored section 5
Fabian Bschorr: Coauthored and edited section 7
Wei Ming Cho: Coauthored section 3.2.1
David Corcos: Wrote section 4
Teresa Guallart-Naval: participated in console and electronics assemblies, coauthored section 2
Heng Jing Han: Coauthored and edited section 5
Kay Chioma Igwe: Coauthored section 7, edited sections 7, 8, and proofread and edited the manuscript
Jacob Kang: Mentored the design team and coauthored section 8

Sebastian Littin: Mentored the gradient team and coauthored sections 4 and 8

Bing Keong (Joe) Li: Mentored the RF team, improved the manuscript

Jie Liu: Contributed to the content of sections 3.1.2 and 7

Gonzalo Gabriel Rodriguez: Participated in magnet assembling and system integration, wrote section 3.1

Eddy Solomon: Coauthored section 1

Li-Kuo Tan: Coauthored section 5

Rui Tian: Wrote section 6

Andrew Webb: Supervised the project, wrote section 7, and edited the manuscript

Susanna Weber: coauthored, reviewed, and edited section 7

Dan Xiao: Coauthored section 1.4

Minxuan Xu: Coauthored section 3.2.1

Wenwei Yu: Mentored the integration team and coauthored section 3

Zhiyong Zhang: Coathored section 7

Paul Cassidy: Coathored section 6

Isabelle Zinghini: Coauthored and reviewed section 4, designed and led fabrication of *Y*-gradient coil, assisted in design and fabrication of *X* and *Y* gradient coils

Bernhard Blümich: Wrote sections 1.1, 1.2, 1.4, 9, coordinated the writing, and edited the manuscript